\documentclass[11pt,a4paper]{article} 
\usepackage[margin=1in]{geometry}
\usepackage{amsmath,amssymb,amsthm,color,enumerate,graphicx,stackrel,natbib,authblk,setspace}
\usepackage{lscape}
\usepackage{placeins}
\usepackage{multirow}

\onehalfspace
\begin{document}

\title{Fluctuation identities with continuous monitoring and their application to price barrier options}

%
%
%
\author{Carolyn E. Phelan

Financial Computing and Analytics Group, Department of Computer Science, University College London

\texttt{c.phelan@cs.ucl.ac.uk}

\bigskip

Daniele Marazzina

Department of Mathematics, Politecnico di Milano

\texttt{daniele.marazzina@polimi.it}

\bigskip

Gianluca Fusai

Department of Economics and Business Studies (DiSEI), Universit\`a del Piemonte Orientale ``Amedeo Avogadro'', Novara\\ Faculty of Finance, Cass Business School, City University London

\texttt{gianluca.fusai@unipmn.it, gianluca.fusai.1@city.ac.uk}

\bigskip

Guido Germano

Financial Computing and Analytics Group, Department of Computer Science, University College London\\ Systemic Risk Centre, London School of Economics

\texttt{g.germano@ucl.ac.uk, g.germano@lse.ac.uk}
}

\maketitle
\begin{abstract}
We present a numerical scheme to calculate fluctuation identities for exponential L\'evy processes in the continuous monitoring case. This includes the Spitzer identities for touching a single upper or lower barrier, and the more difficult case of the two-barriers exit problem. These identities are given in the Fourier-Laplace domain and require numerical inverse transforms. Thus we cover a gap in the literature that has mainly studied the discrete monitoring case; indeed, there are no existing numerical methods that deal with the continuous case.
As a motivating application we price continuously monitored barrier options with the underlying asset modelled by an exponential L\'evy process.
We perform a detailed error analysis of the method and develop error bounds to show how the performance is limited by the truncation error of the sinc-based fast Hilbert transform used for the Wiener-Hopf factorisation. By comparing the results for our new technique with those for the discretely monitored case (which is in the Fourier-$z$ domain) as the monitoring time step approaches zero, we show that the error convergence with continuous monitoring represents a limit for the discretely monitored scheme.
\end{abstract}




\section{Introduction}

Identities providing the Fourier-$z$ transform of probability distribution functions of the extrema of a random path subject to monitoring at discrete intervals were first published by \cite{spitzer1956combinatorial}. They were extended to the continuous case by \cite{baxter1957distribution} and to double barriers by \cite{Kemperman1963}. The identities for the minimum and maximum of a path, for use with a single upper or lower barrier and for the two-barrier exit problem, are comprehensively described in the discrete monitoring case by \cite{Fusai2016}, who proposed numeric methods to compute them for exponential L\'evy processes. The discretely and continuously monitored identities are in the Fourier-$z$ and Fourier-Laplace domains respectively. This means that with the application of the inverse $z$ or Laplace transform as appropriate, they can be used within Fourier-transform option pricing methods, which we will use as an example in this paper. The relevance of the Spitzer identity in several fields within operational research is nowadays well recognised. We mention, for example, the application to queuing systems, see the classical contributions by \cite{cohen1975,cohen1982} and \cite{prabhu1974} and more recent work by \cite{bayer1996}, Markov chains \citep{rogers1994}, insurance \citep{chi2011}, inventory systems \citep{cohen1978,grassman1989}, and applied probability \citep{grassman1990}, as well as in mathematical finance.

Pricing derivatives, especially exotic options, is a challenging problem often covered also in the operations research literature, see e.g.\ \cite{kou2008}. \cite{Fusai2016} provide extensive references for this, as well as for many non-financial applications of the Hilbert transform and the related topics of Wiener-Hopf factorisation and Spitzer identities in insurance, queuing theory, physics, engineering, applied mathematics, etc. Derivative pricing with Fourier transforms was first investigated by \cite{heston1993closed}. \cite{Carr1999} published the first method with both the characteristic function and the payoff in the Fourier domain. \cite{fang2008novel,Fang2009pricing} devised the COS method based on the Fourier-cosine expansion. The Hilbert transform \citep{King2009} has also been successfully employed: by \cite{Feng2008} to price barrier options using backward induction in Fourier space and by \cite{marazzina2012pricing} and \cite{Fusai2016} to compute the factorisations required by the Spitzer identities via the Plemelj-Sokhotsky relations. Feng and Linetsky showed that computing the Hilbert transform with the sinc expansion, as studied by \cite{Stenger1993,Stenger2011}, gives errors that reduce exponentially as the number of fast Fourier transform (FFT) grid points increases. However, the Feng and Linetsky method cannot be extended to continuously monitored options because its recursive structure makes it an inherently discrete scheme. 
In contrast \cite{Green2010} showed that methods based on the Spitzer identities can be extended to continuous monitoring using the Laplace transform in the time domain rather than the $z$-transform.


In this article we implement a method to numerically calculate the required Wiener-Hopf factors and thence the Spitzer identities in continuous time; we apply this to price continuously monitored options with general exponential L\'evy processes.
For continuous monitoring, the Wiener-Hopf factorisation can be done analytically if the characteristic exponent is rational, i.e.\ for the Gaussian and Kou double exponential processes, or in some special cases, e.g.\ when the jumps are only positive or negative. It is also possible to approximate a non-rational exponent with a rational one that is easily factored \citep{kuznetsov2010}. However, an analytical solution for the continuous monitoring case which is usable for any exponential L\'evy process and does not require approximation has not been found yet. In the discrete case an analytical Wiener-Hopf factorisation can be done only for a Gaussian process \citep{fusai2006exact}, but from a numerical point of view the problem is easier and there are a number of papers dealing with exponential L\'evy processes. However, it is well known that the convergence of numerical methods for discrete monitoring to the continuous monitoring limit are very slow, see e.g.~\cite{broadie1997}.
Therefore this work contributes to the literature, both in probability as well as in applied mathematics and operational research, by providing a method to find exit probabilities in the continuous monitoring case with non-rational characteristic exponents, whereas previous numerical methods have concentrated on the discrete monitoring case.

This method follows the approach suggested by \cite{Green2010} and is based on the Fusai, Germano and Marazzina (FGM) method \citep{Fusai2016} with spectral filtering \citep{Phelan2017}. While the latter is for discrete monitoring and thus in the Fourier-$z$ domain, here we operate in the Fourier-Laplace domain. Besides the discrete Fourier transform (DFT), or actually the fast Fourier transform (FFT), which is a standard technique, we also require a numerical inverse Laplace transform; for the latter we used a algorithm proposed by \cite{Abate1992_2,Abate1995}, which is based on a Fourier series and is derived in a similar way to their well established numerical inverse $z$-transform \citep{Abate1992}. The error convergence is slightly worse than first-order polynomial; we explain this in detail with reference to the truncation error of the sinc-based discrete Hilbert transform. Our results show that the error convergence is consistent with the error bound and the performance of the discretely monitored technique as the monitoring interval goes to zero.




The structure of this paper is as follows. In Section 2 we briefly run through Fourier, Hilbert, Laplace and $z$-transforms and explain how they are used for the calculation of the Spitzer identities. We then present a numerical pricing scheme for continuously monitored options and explain its relationship with the FGM pricing scheme with discrete monitoring. Section 3 provides a discussion of the error convergence of the pricing technique with special reference to the truncation error of the sinc-based Hilbert transform. Section 4 shows the results that were achieved, comparing them with the results for the FGM method for discretely monitored options.

\section{Fourier transform methods for option pricing}\label{sec:fourhilb}

In this paper we make extensive use of the Fourier transform \citep[see e.g.][]{Kreyszig2011,Polyanin1998}, an integral transform with many applications. Historically, it has been widely employed in spectroscopy and communications, therefore much of the literature refers to the function in the Fourier domain as its spectrum. According to the usual convention in financial literature, the forward and inverse Fourier transforms are defined as
\begin{align}
\widehat{f}(\xi)&=\mathcal{F}_{x\rightarrow\xi} \left[f(x)\right]=\int^{+\infty}_{-\infty}e^{i\xi x}f(x)dx, \label{eq:FwdFourier}\\
f(x)&=\mathcal{F}^{-1}_{\xi\rightarrow x} \left[\widehat{f}(\xi)\right]=\frac{1}{2\pi}\int^{+\infty}_{-\infty}e^{-i\xi x}\widehat{f}(\xi)d\xi \label{eq:RevFourier}.
\end{align}

Let $S(t)$ be the price of an underlying asset and $x(t) = \log(S(t)/S_0)$ its log-price. To find the price $v(x,t)$ of an option at time $t=0$ when the initial price of the underlying is $S(0) = S_0$, and thus its log-price is $x(0)=0$, we need to discount the expected value of the undamped option payoff $\phi(x(T))e^{-\alpha x(T)}$ at maturity $t=T$ with respect to an appropriate risk-neutral probability distribution function (PDF) $p(x,T)$ whose initial condition is $p(x,0) = \delta(x)$. As shown by
\cite{lewis2001simple}, this can be done using the Plancherel relation,
\begin{align}
v(0,0) & = e^{-rT}\mathrm{E}\left[\phi(x(T))e^{-\alpha x(T)}|x(0)=0\right]=e^{-rT}\int^{+\infty}_{-\infty}\phi(x)e^{-\alpha x}p(x,T)dx \nonumber\\
& = \frac{e^{-rT}}{2\pi}\int^{+\infty}_{-\infty}\widehat{\phi}(\xi)\widehat{p}\,^*(\xi+i\alpha,T)d\xi = e^{-rT}\mathcal{F}^{-1}_{\xi\rightarrow x}\left[\widehat{\phi}(\xi)\widehat{p}\,^*(\xi+i\alpha,T)\right](0). \label{eq:Planch}
\end{align}
Here, $\widehat{p}\,^*(\xi+i\alpha,T)$ is the complex conjugate of the Fourier transform of $e^{-\alpha x}p(x,T)$. To price options using this relation, we need the Fourier transforms of both the damped payoff and the PDF.
A double-barrier option has the damped payoff
\begin{equation}
\label{eq:damped_payoff}
\phi(x) = e^{\alpha x}S_0(\theta(e^x-e^k))^+\mathbf{1}_{[l,u]}(x),
\end{equation}
where $e^{\alpha x}$ is the damping factor, $\theta = 1$ for a call, $\theta = -1$ for a put, $\mathbf{1}_A(x)$ is the indicator function of the set $A$, $k=\log(K/S_0)$ is the log-strike, $u=\log(U/S_0)$ is the upper log-barrier, $l=\log(L/S_0)$ is the lower log-barrier, $K$ is the strike price, $U$ is the upper barrier and $L$ is the lower barrier. The Fourier transform of the damped payoff $\phi(x)$ is available analytically, 
\begin{equation}
\label{eq:Payoff}
\widehat{\phi}(\xi)=S_0\left(\frac{e^{(1+i\xi+\alpha)a}-e^{(1+i\xi+\alpha)b}}{1+i\xi+\alpha}-\frac{e^{k+(i\xi+\alpha)a}-e^{k+(i\xi+\alpha)b}}{i\xi+\alpha}\right),
\end{equation}
where for a call option $a = u$ and $b = \max(k,l)$, while for a put option $a=l$ and $b=\min(k,u)$.

The Fourier transform of the PDF $p(x,t)$ of a stochastic process $X(t)$ is the characteristic function
\begin{equation}
\label{eq:CharFun}
\Psi(\xi,t)=\mathrm{E}\left[e^{i\xi X(t)}\right]=\int^{+\infty}_{-\infty}e^{i\xi x}p(x,t)dx=\mathcal{F}_{x\rightarrow\xi}\left[p(x,t)\right]=\widehat{p}(\xi,t).
\end{equation}
For a L\'evy process the characteristic function can be written as $\Psi(\xi,t)=e^{\psi(\xi)t}$, where the characteristic exponent $\psi(\xi)$ is given by the L\'evy-Khincine formula as
\begin{equation}
\label{eq:CharExp}
\psi(\xi)=i\mu\xi-\frac{1}{2}\sigma^2\xi^2+\int_{\mathbb{R}}(e^{i\xi\eta}-1-i\xi\eta\mathbf{1}_{[-1,1]}(\eta))\nu(d\eta).
\end{equation}
The L\'evy-Khincine triplet $(\mu,\sigma,\nu)$ uniquely defines the L\'evy process: $\mu$ defines the linear drift of the process, $\sigma$ is the volatility of the diffusion part of the process, and the jump part of the process is specified so that $\nu(\eta)$ is the intensity of a Poisson process with jump size $\eta$. Under the risk-neutral measure the parameters of the triplet are linked by the equation
\begin{equation}
\mu = r-q-\frac{1}{2}\sigma^2-\int_\mathbb{R}(e^\eta-1-i\eta\mathbf{1}_{[-1,1]}(\eta))\nu(d\eta),
\end{equation}
where $r$ is the risk-free interest rate and $q$ is the dividend rate.
In general the characteristic function of a L\'evy process is available in closed form, for example for the Gaussian \citep{Schoutens2003}, normal inverse Gaussian (NIG) \citep{Barndorff1998}, CGMY \citep{Carr2002}, Kou double exponential \citep{kou2002jump}, Merton jump diffusion \citep{merton1976option}, L\'evy alpha stable \citep{Nolan2017}, Variance Gamma (VG) \citep{Madan1990} and Meixner \citep{Schoutens2003} processes.

Some pricing techniques based on the Fourier transform also use the Hilbert transform, which is an integral transform related to the Fourier transform. Unlike with the Fourier transform, the function under transformation remains in the same domain, rather than moving between the $x$ and $\xi$ domains. The Hilbert transform of a function in the Fourier domain is defined as
\begin{align}
\label{eq:HilbTrans}
\mathcal{H}\big[\widehat{f}(\xi)\big]&=\,\mathrm{P.V.}\,\frac{1}{\pi}\int^{+\infty}_{-\infty}\frac{\widehat{f}(\xi')}{\xi-\xi'}d\xi'\nonumber\\
&=\lim_{\epsilon\rightarrow0^+}\frac{1}{\pi}\left(\int_{\xi-1/\epsilon}^{\xi-\epsilon} \frac{\widehat{f}(\xi')}{\xi-\xi'} d\xi'+\int_{\xi+\epsilon}^{\xi+1/\epsilon} \frac{\widehat{f}(\xi')}{\xi-\xi'} d\xi'\right),
\end{align}
where $\mathrm{P.V.}$ denotes the Cauchy principal value. Applying the Hilbert transform in the Fourier domain is equivalent to multiplying the function in the $x$ domain by $-i\,\mathrm{sgn}\,x$.

Whilst the Fourier and Hilbert transform operate on the state variable (here the log price), the Laplace transform is applied to time. The forward and reverse Laplace transforms are
\begin{align}
\mathcal{L}_{t\rightarrow s}[f(t)]&=\widetilde{f}(s):=\int_{0}^{+\infty}e^{-st}f(t)dt\label{eq:fwdLap}, \quad s\in\mathbb{C}\\
\mathcal{L}_{s\rightarrow t}^{-1}[\widetilde{f}(s)]&= f(t) := \frac{1}{2\pi i}\int^{a+i\infty}_{a-i\infty} e^{st}\widetilde{f}(s)ds,\label{eq:invLap}
\end{align}
where $a\in\mathbb{R}$ is on the right of all singularities of $\widetilde{f}(s)$ in the complex plane. The Laplace transform is closely related to the $z$-transform of a discrete function $f(t_n) = f(n)$, $n\in\mathbb{N}_0$,
\begin{equation}
\label{eq:ZFor}
\mathcal{Z}_{n\rightarrow q}[f(n)]:=\sum_{n=0}^\infty q^n f(t_n),\quad q\in\mathbb{C}.
\end{equation}
Given a continuous function $f_{\mathrm{c}}(t)$, we define the discrete function $f_{\mathrm{d}}(t_n)$ consisting of sampled values of the former, where $\Delta t$ is the sampling interval and $t_n=n\Delta t$ are the sampling times. Then with a $z$-transform parameter $q=e^{-s\Delta t}$, the Laplace and $z$-transforms are related in the limit $\Delta t\rightarrow\infty$:
\begin{align}
\label{eq:ztolap}
\mathcal{L}_{t\rightarrow s}[f_{\mathrm{c}}(t)]&=\int_0^\infty e^{-st}f_{\mathrm{c}}(t)dt
=\lim_{\Delta t\rightarrow0}\Delta t\sum_{n=0}^\infty e^{-sn\Delta t} f_\mathrm{c}(n\Delta t) \nonumber \\
& =\lim_{\Delta t\rightarrow0}\Delta t\sum_{n=0}^\infty (e^{-s\Delta t})^n f_\mathrm{d}(t_n)
=\lim_{\Delta t\rightarrow0}\Delta t\sum_{n=0}^\infty q^nf_\mathrm{d}(t_n) \nonumber\\
& =\lim_{\Delta t\rightarrow0}\Delta t \mathcal{Z}\left[f_\mathrm{d}(t_n)\right].
\end{align}

\subsection{Spitzer identities for continuous monitoring}\label{sec:back_spitzer}

If we wish to use Eq.~(\ref{eq:Planch}) to price barrier options, the required characteristic functions are more complicated than the closed-form expressions referred to above. We need the characteristic function of the PDF of the value of a stochastic process $X(t)$ at time $t=T$, conditional on the process remaining inside continuously monitored upper and lower barriers. We use the identities published by \cite{spitzer1956combinatorial} which were extended to the continuously monitored case by \cite{baxter1957distribution} and to double-barriers by \cite{Kemperman1963}. The Spitzer identities provide the Fourier-$z$ transform of the PDF of a stochastic process $X(t)$ at time $t=T$, conditional on whether $X(t)$ reaches a barrier at discretely monitored times. The Fourier transform is applied to the process values and the $z$-transform is applied to the discrete monitoring times.
\cite{baxter1957distribution} demonstrated that we can obtain the equivalent identities for continuously monitored barriers in the Fourier-Laplace domain. \cite{Green2010} showed that the relationship between the Laplace and $z$-transforms described in Eq.~(\ref{eq:ztolap}) can be exploited to price continuously monitored options using the Spitzer identities in the Fourier-Laplace domain.

An important aspect in the calculation of the Spitzer identities is the decomposition of a function $\widehat{f}(\xi)$ into $+$ and $-$ parts, $\widehat{f_+}(\xi)=\mathcal{F}_{x\rightarrow\xi}\left[f(x)\mathbf{1}_{\mathbb{R}_+}(x)\right]$ and $\widehat{f_-}(\xi)=\mathcal{F}_{x\rightarrow\xi}\left[f(x)\mathbf{1}_{\mathbb{R}_-}(x)\right]$, such that $\widehat{f}(\xi) = \widehat{f_+}(\xi) + \widehat{f_-}(\xi)$.
This can be done directly in the Fourier domain using the Plemelj-Sokhotsky relations \citep{King2009,Fusai2016}:
\begin{align}
\label{eq:PSRel}
\widehat{f_+}(\xi)&=\frac{1}{2}\big\{\widehat{f}(\xi)+i\mathcal{H}\big[\widehat{f}(\xi)\big]\big\}\\
\widehat{f_-}(\xi)&=\frac{1}{2}\big\{\widehat{f}(\xi)-i\mathcal{H}\big[\widehat{f}(\xi)\big]\big\}.
\end{align}
The shift theorem $\mathcal{F}_{x\rightarrow\xi}[f(x+b)]=\widehat{f}(\xi)e^{-ib\xi}$ allows to obtain the generalised Plemelj-Sokhotsky relations for an arbitrary barrier $b$:
\begin{align}
\widehat{f_{b+}}(\xi)&=\frac{1}{2}\big\{\widehat{f}(\xi)+e^{ib\xi}i\mathcal{H}\big[e^{-ib\xi}\widehat{f}(\xi)\big]\big\}\label{eq:PSRelgenpos}\\
\widehat{f_{b-}}(\xi)&=\frac{1}{2}\big\{\widehat{f}(\xi)-e^{ib\xi}i\mathcal{H}\big[e^{-ib\xi}\widehat{f}(\xi)\big]\big\}.\label{eq:PSRelgenneg}
\end{align}
The calculation of the Spitzer identities also requires to factorise a function, i.e.~obtain $\widehat{g_+}(\xi)$ and $\widehat{g_-}(\xi)$ such that $\widehat{g}(\xi)=\widehat{g_+}(\xi)\widehat{g_-}(\xi)$. This is achieved by decomposing the logarithm $\widehat{h}(\xi)=\log\widehat{g}(\xi)$ and then exponentiating the results to obtain $\widehat{g_+}(\xi)=\exp\widehat{h_+}(\xi)$ and $\widehat{g_-}(\xi)=\exp\widehat{h_-}(\xi)$.

\cite{Green2010} dealt with fluctuation identities that can be used for lookback, single-barrier and double-barrier options. Here we concentrate on the identities for single-barrier down-and-out and double-barrier options. The first step is always to factorise $\Phi_\mathrm{c}(\xi,s)=s-\psi(\xi)=\Phi_{\mathrm{c}+}(\xi,s)\Phi_{\mathrm{c}-}(\xi,s)$. For a single-barrier down-and-out option, the Laplace transform of the required characteristic function is
\begin{equation}
\label{eq:Pcdecomp1}
\widetilde{\widehat{p}}(\xi,s) =\frac{1-\Phi_{\mathrm{c}-}(\xi,s)P_{\mathrm{c}l-}(\xi,s)}{\Phi_{\mathrm{c}}(\xi,s)} =\frac{P_{\mathrm{c}l+}(\xi,s)}{\Phi_{\mathrm{c}+}(\xi,s)},
\end{equation}
where $P_{\mathrm{c}}(\xi,s) = 1/\Phi_{\mathrm{c}-}(\xi,s)$ is decomposed with respect to the lower log-barrier $l$ using Eqs.~(\ref{eq:PSRelgenpos}) and (\ref{eq:PSRelgenneg}).
For a double-barrier option, the Laplace transform of the required characteristic function is
\begin{equation}
\label{eq:Doubcpdf}
\widetilde{\widehat{p}}(\xi,s)=\frac{1-\Phi_{\mathrm{c}-}(\xi,s)J_{\mathrm{c}l-}(\xi,s)-\Phi_{\mathrm{c}+}(\xi,s)J_{\mathrm{c}u+}(\xi,s)}{\Phi_\mathrm{c}(\xi,s)},
\end{equation}
where $J_{\mathrm{c}u+}(\xi,s)$ and $J_{\mathrm{c}l-}(\xi,s)$ are the solution to the pair of coupled equations
\begin{align}
J_{\mathrm{c}u+}(\xi,s)&=\left[\frac{1-\Phi_{\mathrm{c}-}(\xi,s)J_{\mathrm{c}l-}(\xi,s)}{\Phi_{\mathrm{c}+}(\xi,s)}\right]_{u+}\label{eq:Jmin} \\
J_{\mathrm{c}l-}(\xi,s)&=\left[\frac{1-\Phi_{\mathrm{c}+}(\xi,s)J_{\mathrm{c}u+}(\xi,s)}{\Phi_{\mathrm{c}-}(\xi,s)}\right]_{l-}.\label{eq:Jpos}
\end{align}
\noindent For $u\to\infty$, $J_{\mathrm{c}u+}\to 0$ and $J_{\mathrm{c}l-}\to P_{\mathrm{c}l-}$, thus recovering the Spitzer identity for the single barrier, Eq.~(\ref{eq:Pcdecomp1}). The latter can be calculated directly, while so far only an iterative solution has been found \citep{Fusai2016,Phelan2017} to the coupled Eqs.~(\ref{eq:Jmin}) and (\ref{eq:Jpos}).

\subsubsection{Relationship to the Spitzer identities for discrete monitoring}\label{sec:back_spitz_rel}

In Section \ref{sec:Res} we show numerical results comparing the error convergence obtained using the Spitzer identities for continuous monitoring with the performance of the closely related method using the Spitzer identities for discrete monitoring \citep{Green2010,Fusai2016,Phelan2017}.

The relationship between the two methods originates in the connection between the $z$-transform and the Laplace transform described in Eq.~(\ref{eq:ztolap}). As described in Section \ref{sec:back_spitzer}, the first step in pricing continuously monitored barrier options is the calculation of $\Phi_\mathrm{c}(\xi,s)=s-\psi(\xi)$ in the Fourier-Laplace domain. The equivalent quantity in the Fourier-$z$ domain for discrete monitoring is $\Phi(\xi,q)=1-q\Psi(\xi,\Delta t)$.
We can use the relation in Eq.~(\ref{eq:ztolap}) with $q=e^{-s\Delta t}$ to relate the two:
\begin{align}
\label{eq:LapPhi}
\lim_{\Delta t\rightarrow0}\frac{\Delta t}{\Phi(\xi,q)}& = \lim_{\Delta t\rightarrow0}\frac{\Delta t}{1-q\Psi(\xi,\Delta t)}
= \lim_{\Delta t\rightarrow0}\frac{\Delta t}{1-e^{-s\Delta t}e^{\psi(\xi)\Delta t}}\nonumber\\
& = \lim_{\Delta t\rightarrow0}\frac{\Delta t}{1-e^{(\psi(\xi)-s)\Delta t}}
= \frac{1}{s-\psi(\xi)}
= \frac{1}{\Phi_{\mathrm{c}}(\xi,s)}.
\end{align}
The same factorisation and decomposition steps described in Section \ref{sec:back_spitzer} \citep{Green2010,Fusai2016} are applied to both $\Phi(\xi,q)$ and $\Phi_{\mathrm{c}}(\xi,s)$ to price options with respectively discrete or continuous monitoring.

\subsection{Numerical methods}

The methods in the previous section are described analytically. However, as they involve some expressions which cannot be solved in closed form, their implementation requires the use of numerical approximation techniques which we discuss in the following sections.

\subsubsection{Discrete Fourier and Hilbert transforms and spectral filters}

The forward and inverse Fourier transforms shown in Eq.~(\ref{eq:FwdFourier}) and Eq.~(\ref{eq:RevFourier}) are integrals over an infinite domain and in order to implement them numerically we need to approximate each with a discrete Fourier transform (DFT).
We implement this in practice using the built-in Matlab FFT function which is based on the FFTW library by \cite{frigo1998fftw}.
\FloatBarrier

The calculation of the Hilbert transform of a function $\widehat{f}(\xi)$ can be realised with an inverse/forward Fourier transform pair and multiplication by the sign function in between,
\begin{equation}
i\mathcal{H}\big[\widehat{f}(\xi)\big]=\mathcal{F}_{x\rightarrow\xi}\big[\mathrm{sgn}(x)\mathcal{F}^{-1}_{\xi\rightarrow x}\widehat{f}(\xi)\big].
\end{equation}
However, this gives an error convergence which is polynomially decreasing with the number of grid points $M$. In order to obtain exponential error convergence, \cite{Feng2008} and \cite{Fusai2016} have implemented the Hilbert transform using the sinc expansion techniques studied by \cite{Stenger1993,Stenger2011}. Stenger showed that, given a function $\widehat{f}(\xi)$ which is analytic in the whole plane, the function and its Hilbert transform can be expressed as
\begin{align}
\widehat{f}(\xi)&=\sum^{+\infty}_{k=-\infty}\widehat{f}(k\Delta\xi)\frac{\sin(\pi(\xi-k\Delta\xi)/\Delta\xi)}{\pi(\xi-k\Delta\xi)/\Delta\xi},\label{eq:SincApprox}\\
\mathcal{H}\big[\widehat{f}(\xi)\big]&=\sum^{+\infty}_{k=-\infty}\widehat{f}(k\Delta\xi)\frac{1-\cos(\pi(\xi-k\Delta\xi)/\Delta\xi)}{\pi(\xi-k\Delta\xi)/\Delta\xi}, \label{eq:HilbSincApprox}
\end{align}
where $\Delta\xi$ is the grid step in the Fourier domain. \cite{Stenger1993} also showed that, when the function $f(\xi)$ is analytic in a strip of the complex plane including the real axis, the expressions in Eqs.~(\ref{eq:SincApprox}) and (\ref{eq:HilbSincApprox}) are approximations whose error decays exponentially as $\Delta\xi$ decreases. In addition to discretisation, the infinite sum in Eq.~(\ref{eq:HilbSincApprox}) must also be truncated to the grid size $M$, so that the Hilbert transform approximation becomes
\begin{equation}
\label{eq:HilbSincApproxTrunc}
\mathcal{H}\big[\widehat{f}(\xi)\big]\approx\sum^{+M/2}_{k=-M/2}\widehat{f}(k\Delta\xi)\frac{1-\cos(\pi(\xi-k\Delta\xi)/\Delta\xi)}{\pi(\xi-k\Delta\xi)/\Delta\xi}.
\end{equation}
\cite{Feng2008,Feng2009} showed that if $\widehat{f}(\xi)$ decays at least exponentially as $|\xi|\rightarrow\infty$, i.e.~$\widehat{f}(\xi)\leq \kappa \exp(-c |\xi|^{\nu})$, then the error in the Hilbert transform and thus in the Plemelj-Sokhotsky relations caused by truncating the series in Eq.~(\ref{eq:HilbSincApprox}) is also exponentially bounded. Furthermore Feng and Linetsky showed that if $\widehat{f}(\xi)$ is polynomially bounded then, although the accuracy of the series in Eq.~(\ref{eq:HilbSincApprox}) is retained, the error caused by truncating the sum is no longer exponentially bounded. 
However, it has subsequently been shown that multiplying the input to the Hilbert transform by a filter can improve the error convergence \citep{Phelan2017}.

\begin{figure}
\begin{center}
\includegraphics[width=\textwidth]{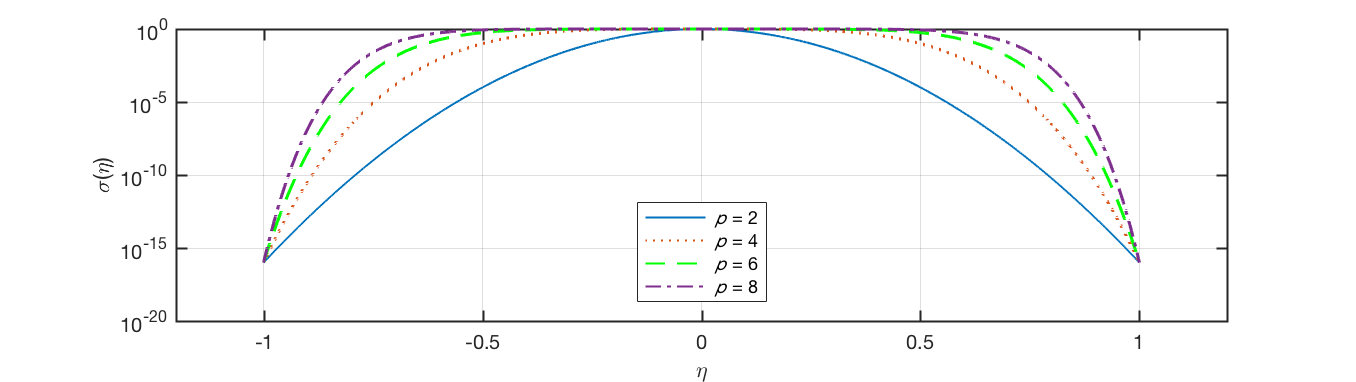}
\caption{Shape of the exponential filter plotted with different values of $p$.}
\label{fig:expfilt2}
\end{center}
\end{figure}

In the papers by \cite{gottlieb1997gibbs} and \cite{vandeven1991family}, a filter of order $p$ is defined as a function $\sigma(\eta)$ supported on $\eta\in[-1,1]$ with the properties
\begin{align}
\label{eq:filtDef}
& \text{a) }\sigma(0)=1\text{,    } \sigma^{(l)}(0)=0 \nonumber\\
& \text{b) }\sigma(\eta)=0 \text{ for }|\eta|=1\\
& \text{c) }\sigma(\eta)\in C^{p-1}.\nonumber
\end{align}
The scaled variable $\eta$ is related to $\xi$ in our application as $\eta=\xi/\xi_{\max}$. In this paper we use the exponential filter, which has the form \citep{gottlieb1997gibbs}
\begin{equation}
\label{eq:expFilt}
\sigma(\eta)=e^{-\vartheta\eta^p},
\end{equation}
where $p$ is even. This does not strictly meet criterion b in Eq.~(\ref{eq:filtDef}) as it does not go exactly to zero when $|\eta|=1$. However, if we select $\vartheta<\varepsilon\log 10$, where $10^{-\varepsilon}$ is the machine precision, then the filter coefficients are within computational accuracy of the requirements. The exponential filter has the advantages that it has a simple form and that it can be used for any even value of $p$. Moreover, the order of the filter is a parameter which is directly input to the filter equation. Filter shapes for a range of $p$ values are shown in Figure \ref{fig:expfilt2}.
Many filters other than the exponential can be used, e.g.\ the Planck taper \citep{Phelan2017} and the raised cosine \citep{ruijter2015application}.

\subsubsection{Inverse Laplace transform}

\begin{figure}
\begin{center}
\includegraphics[width=\textwidth]{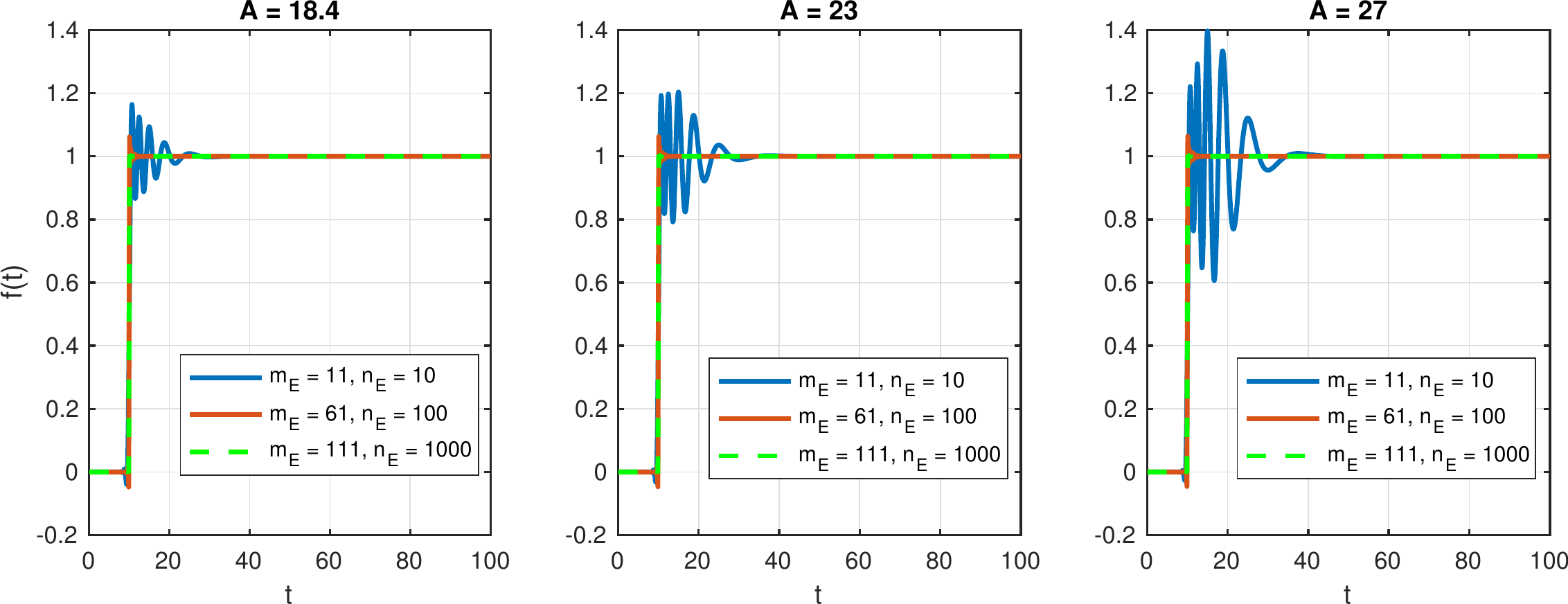}
\caption{Output of the inverse Laplace transform of $\widetilde{f}(s)=\frac{e^{-10s}}{s}$. Increasing $n_\mathrm{E}$ and $m_\mathrm{E}$ reduces the size of the oscillations, but it is not improved by increasing $A$.}
\label{fig:invlapstep}
\end{center}
\end{figure}

\begin{figure}
\begin{center}
\includegraphics[width=\textwidth]{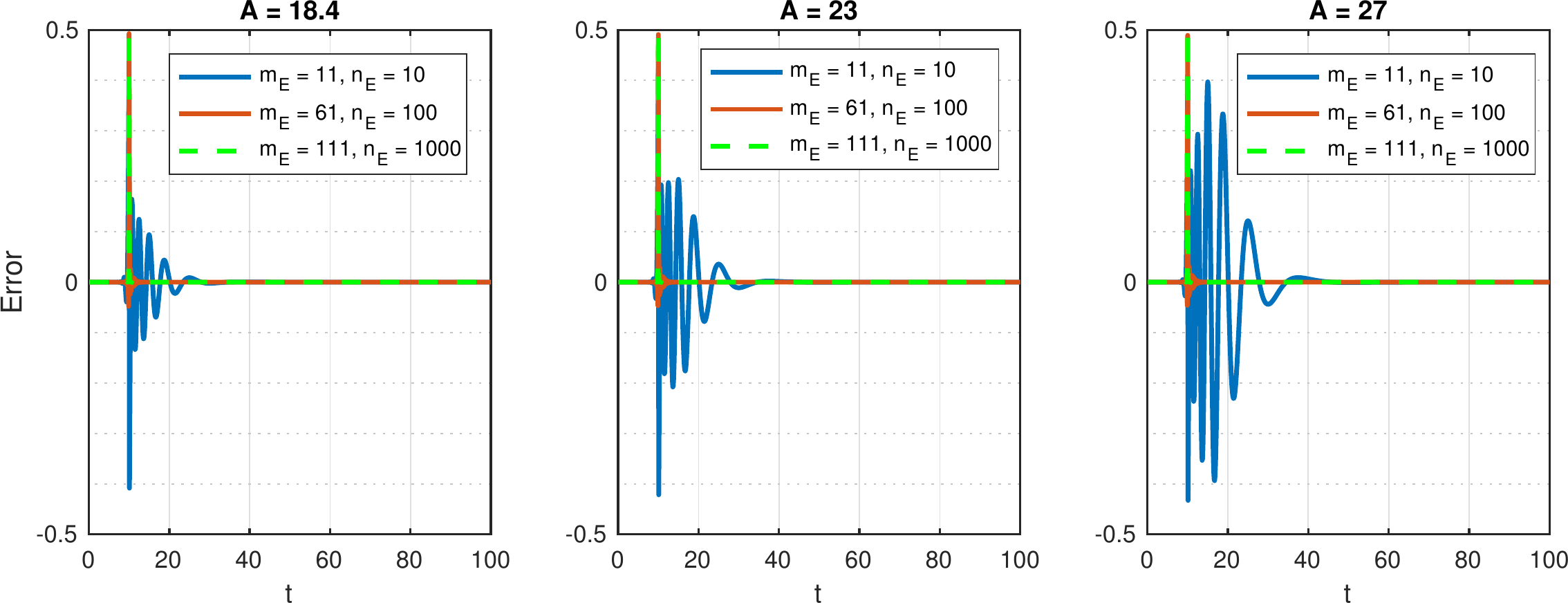}
\caption{Error of the inverse Laplace transform of $\widetilde{f}(s)=\frac{e^{-10s}}{s}$. Increasing $n_\mathrm{E}$ and $m_\mathrm{E}$ reduces the size of the errors due to the oscillations, but it is not improved by increasing $A$.}
\label{fig:invlaperr1}
\end{center}
\end{figure}

\begin{figure}
\begin{center}
\includegraphics[width=\textwidth]{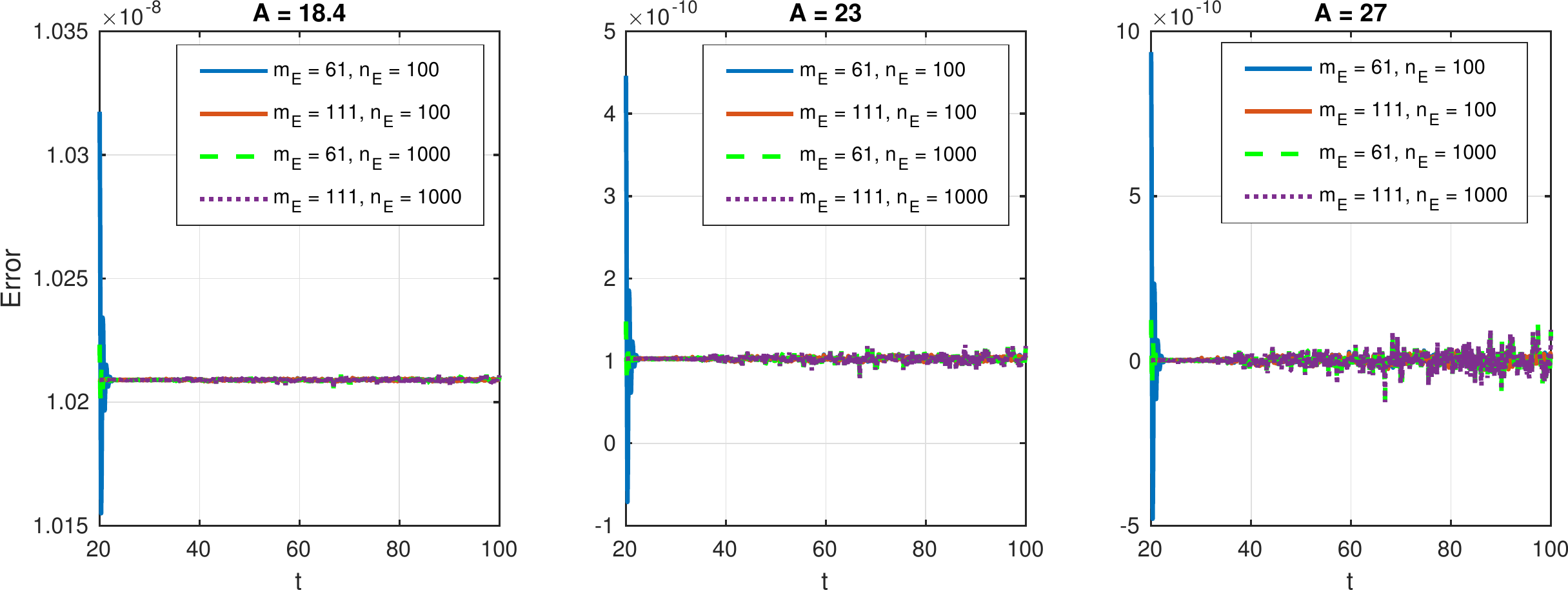}
\caption{Error of the inverse Laplace transform of $\widetilde{f}(s)=\frac{e^{-10s}}{s}$. Increasing $A$ decreases the error floor, while the latter is unaffected by increasing $n_\mathrm{E}$ and $m_\mathrm{E}$. The noise on the error floor is $\approx10^{-10}$.}
\label{fig:invlaperr2}
\end{center}
\end{figure}

The Spitzer identities provide the Laplace transform of the characteristic function, so to calculate the option price using Eq.~(\ref{eq:Planch}) we must apply the inverse Laplace transform. We implement the numerical scheme by \cite{Abate1995}, which uses the trapezoidal rule to approximate the analytic expression for the inverse Laplace transform shown in Eq.~(\ref{eq:invLap}) with
\begin{equation}
f(t) \approx \frac{e^{A/2}}{2t}\mathrm{Re}\widetilde{f}\left(\frac{A}{2t}\right)+\frac{e^{A/2}}{t}\sum_{k=1}^{\infty}(-1)^{k}\mathrm{Re}\widetilde{f}\left(\frac{A+2k\pi i}{2t}\right),\label{eq:invLapft}
\end{equation}
where $\widetilde{f}\left(\frac{A+2k\pi i}{2t}\right)$ is the Laplace transform $\widetilde{f}(s)$ with $s=\frac{A+2k\pi i}{2t}$. The value of $A$ is selected to control the accuracy of the approximation; for an accuracy of $10^{-\gamma}$ we must select $A=\gamma \log(10)$.
We then use the Euler transform to accurately approximate this infinite series. First the partial sums
\begin{equation}
b_k = \frac{e^{A/2}}{2t}\mathrm{Re}\widetilde{f}\left(\frac{A}{2t}\right)+\frac{e^{A/2}}{t}\sum_{j=1}^{k}(-1)^{j}\mathrm{Re}\widetilde{f}\left(\frac{A+2j\pi i}{2t}\right)\label{eq:invLapSl}
\end{equation}
are calculated for $k=n_\mathrm{E},\dots,n_\mathrm{E}+m_\mathrm{E}$. We then take the binomially weighted average (Euler transform) of these terms, i.e.,
\begin{equation}
f(t) \approx \frac{1}{2^{m_\mathrm{E}}}\sum_{k=0}^{m_\mathrm{E}}\binom{m_\mathrm{E}}{k}b_{n_\mathrm{E}+k}.
\end{equation}
The values of $n_\mathrm{E}$ and $m_\mathrm{E}$ are selected large enough to give sufficient accuracy, but low enough to avoid unnecessary computational effort.
Numerical tests were carried out inverting the Laplace transform of a delayed unit step function $\widetilde{f}(s)=e^{-\tau s}/s$ where the delay $\tau = 10$. This is an extreme test case as the step function has a jump discontinuity and \cite{Abate1992_2} state that the performance bound of $10^{-\gamma}=e^{-A}$ does not apply in the presence of jumps. However it is important to consider the performance of the inverse Laplace transform with discontinuities in the time domain as the value of the contracts that we are pricing will abruptly become zero on expiry. The recovered functions for different values of $A$, $n_\mathrm{E}$ and $m_\mathrm{E}$ are shown in Figure \ref{fig:invlapstep} and the errors are shown in Figures \ref{fig:invlaperr1} and \ref{fig:invlaperr2}. The empirical results in Figure \ref{fig:invlaperr2} show that we can select values for $A$, $m_\mathrm{E}$ and $n_\mathrm{E}$ so that, away from the discontinuity, the performance matches the bound of $10^{-\gamma}=e^{-A}$ specified by Abate and Whitt. Furthermore, we show in Sections \ref{sec:errperf} and \ref{sec:Res} that the error bounds and observed results for the pricing procedure are limited by the performance of the sinc-based Hilbert transform. Therefore, we can use the Abate and Whitt inverse Laplace transform method to price mid- to long-dated options.

We base the selection of the parameters for the inverse Laplace transform on the empirical results. From Figures \ref{fig:invlapstep} and \ref{fig:invlaperr1} we can see that the size of the oscillations due to the discontinuity are predominantly affected by $m_\mathrm{E}$ and $n_\mathrm{E}$. The error floor is controlled by $A$; the values of 18.4, 23 and 27 in Figures \ref{fig:invlapstep}--\ref{fig:invlaperr2} correspond to errors of approximately $10^{-8}$, $10^{-10}$ and $10^{-12}$ respectively. However, Figure \ref{fig:invlaperr2} shows that the noise around the error floor is $\approx10^{-10}$ and therefore there is no advantage in selecting values of $A$ larger than $23$. For the pricing calculations we use $A=23$, $m_\mathrm{E}=61$ and $n_\mathrm{E}=100$ which give a combination of high accuracy and fast computation time.

\subsubsection{Pricing procedure: single-barrier options}\label{sec:updatedproc_sing}

We describe the pricing procedure for single-barrier down-and-out options as an example, but the use of the Spitzer identities is equally applicable to other types of barrier options and also to lookback options; the pricing formulae described by \cite{Green2010} include methods for single-barrier up-and-out and knock-in options. The pricing method is adapted from the scheme by \cite{Fusai2016} and \cite{Phelan2017} using the relationship between $\Phi(\xi,q)$ and $\Phi_{\mathrm{c}}(\xi,s)$ described in Section \ref{sec:back_spitz_rel}.
\begin{enumerate}
\item Compute the characteristic exponent $\psi(\xi+i\alpha)$, where $\alpha$ is the damping parameter introduced in Section \ref{sec:fourhilb}, Eq.~(\ref{eq:damped_payoff}).
\item Use the Plemelj-Sokhotsky relations with the sinc-based Hilbert transform to factorise
\begin{equation}
\label{eq:Phifact}
\Phi_{\mathrm{c}}(\xi,s):=s-\psi(\xi+i\alpha)=\Phi_{\mathrm{c}+}(\xi,s)\Phi_{\mathrm{c}-}(\xi,s)
\end{equation}
for all $s=\frac{A+2k\pi i}{2t}$ required for the inverse Laplace transform in Eq.~(\ref{eq:invLapSl}).
\item Decompose with respect to $l$ \label{item:Pdecomp}
\begin{equation}
\label{eq:Pdecomp1}
P_{\mathrm{c}}(\xi,s) := \frac{\sigma(\xi/\xi_{\max})}{\Phi_{\mathrm{c}-}(\xi,s)} = P_{\mathrm{c}l+}(\xi,s)+P_{\mathrm{c}l-}(\xi,s),
\end{equation}
and calculate
\begin{equation}
\label{eq:Probdist}
\widetilde{\widehat{p}}(\xi,s) := \frac{P_{\mathrm{c}l+}(\xi,s)}{\Phi_{\mathrm{c+}}(\xi,s)},
\end{equation}
where $\sigma(\xi/\xi_{\max})$ is an exponential filter of order $p$ (see Section \ref{sec:errperf_decomp}).
\item Calculate the option price as
\begin{equation}
\label{eq:Price}
v(0,0) := \mathcal{F}^{-1}_{\xi\rightarrow x}\left[\widehat{\phi}^*(\xi)\mathcal{L}^{-1}_{s\rightarrow T}\widetilde{\widehat{p}}(\xi,s)\right](0),
\end{equation}
where $\widehat{\phi}^*(\xi)$ is the complex conjugate of the Fourier transform of the damped payoff function given in Eq.~(\ref{eq:Payoff}).
\end{enumerate}

\subsubsection{Pricing procedure: double-barrier options}\label{sec:updatedproc_doub}

The pricing procedure for double-barrier options is very similar to the method for the single-barrier options described in Section \ref{sec:updatedproc_sing}, in that it uses Wiener-Hopf factorisation and decomposition to compute the appropriate Spitzer identity. However, the major difference in this case is that the equations cannot be solved directly and so require the use of a fixed point algorithm. The steps in the pricing procedure are the same as those for single-barrier down-and-out options described in Section \ref{sec:updatedproc_sing} with the exception of Step 3 which is now replaced by the fixed-point algorithm
\begin{enumerate}
\item[3 (a)] Set $J_{\mathrm{c}u+}(\xi,s)=J_{\mathrm{c}l-}(\xi,s)=0$.
\item[\phantom{3. }(b)] Decompose with respect to $l$
\begin{equation}
\label{eq:Pcdecomp}
P_{\mathrm{c}}(\xi,s):=\sigma\left(\frac{\xi}{\xi_{\max}}\right)\frac{1-\Phi_{\mathrm{c}+}(\xi,s)J_{\mathrm{c}u+}(\xi,s)}{\Phi_{\mathrm{c}-}(\xi,s)} = P_{\mathrm{c}l+}(\xi,s)+P_{\mathrm{c}l-}(\xi,s),
\end{equation}
and set $J_{\mathrm{c}l-}(\xi,s):=P_{\mathrm{c}l-}(\xi,s)$.
\item[\phantom{3. }(c)] Decompose with respect to $u$
\begin{equation}
\label{eq:Qdecomp}
Q_{\mathrm{c}}(\xi,s):=\sigma\left(\frac{\xi}{\xi_{\max}}\right)\frac{1-\Phi_{\mathrm{c}-}(\xi,s)J_{\mathrm{c}l-}(\xi,s)}{\Phi_{\mathrm{c}+}(\xi,s)} = Q_{\mathrm{c}u+}(\xi,s)+Q_{\mathrm{c}u-}(\xi,s),
\end{equation}
and set $J_{\mathrm{c}u+}(\xi,s):=Q_{\mathrm{c}u+}(\xi,s)$.
\item[\phantom{3. }(d)] Calculate
\begin{equation}
\label{eq:FinalfpCalc}
\widetilde{\widehat{p}}(\xi,s):=\sigma\left(\frac{\xi}{\xi_{\max}}\right)\frac{1-\Phi_{\mathrm{c}-}(\xi,s)J_{\mathrm{c}l-}(\xi,s)-\Phi_{\mathrm{c}+}(\xi,s)J_{\mathrm{c}u+}(\xi,s)}{\Phi_\mathrm{c}(\xi,s)}.
\end{equation}
\item[\phantom{3. }(e)] If the difference between the new and the old value of $\widetilde{\widehat{p}}(\xi,s)$ is less than a predefined tolerance or the number of iterations is greater than a certain threshold then continue, otherwise return to step (b). Numerical tests have shown that an iteration threshold of 5 is sufficient, as higher values do not yield improvements.
\end{enumerate}

\section{Error convergence of the pricing procedure}\label{sec:errperf}

We examine the performance of each stage of the pricing procedure and discuss the respective error bounds. In addition, the effect of each step on the shape of the output function in the Fourier domain is investigated, as this influences the error convergence of later steps. \cite{Stenger1993} showed that the discretisation error in Eq.~(\ref{eq:HilbSincApprox}) is exponentially convergent when the function $f(\xi)$ is analytic in a strip of the complex plane including the real axis. Therefore the error calculations here concern the truncation error from the approximation in Eq.~(\ref{eq:HilbSincApproxTrunc}). The truncation error using the sinc-based Hilbert transform depends on the behaviour of the characteristic exponent as $|\xi|\rightarrow\infty$: Table \ref{tab:Charexp} shows the characteristic exponents of five L\'evy processes. The damping parameter $\alpha$ is omitted to make the notation more concise, which is appropriate as its value becomes insignificant as $|\xi|\rightarrow\infty$.

\begin{table}[h]
\renewcommand{\arraystretch}{1.5}
\centering
\begin{tabular}{ll}
\hline\hline
Process & Characteristic exponent $\psi(\xi)$ \\
\hline
Normal & $i\xi\mu-\frac{1}{2}\sigma^{2}\xi^{2}$ \\
Kou & $i\xi\mu-\frac{1}{2}\sigma^{2}\xi^{2}+\lambda\left(\frac{(1-\rho)\eta_{2}}{\eta_{2}+i\xi}+\frac{\rho\eta_{1}}{\eta_{1}-i\xi}\right)$ \\
Merton & $i\xi\mu-\frac{1}{2}\sigma^{2}\xi^{2}+\lambda\left(e^{i\alpha\xi-\frac{1}{2}\delta^{2}\xi^{2}}-1\right)$ \\
NIG & $\delta\left(\sqrt{\alpha^{2}-(\beta+i\xi)^{2}}-\sqrt{\alpha^{2}-\beta^{2}}\right)$ \\
VG & $-\frac{1}{\nu}\log\left(1-i\xi\theta\nu+\frac{1}{2}\nu\sigma^2\xi^2\right)$ \\
\hline\hline
\end{tabular}
\caption{Characteristic exponent of some L\'evy processes.}
\label{tab:Charexp}
\end{table}

\subsection{Factorisation}

After calculating the characteristic exponent, the next step in the pricing procedure is the numerical factorisation of $\Phi_{\mathrm{c}}(\xi,s)=s-\psi(\xi)$. In order to understand the error convergence we must consider the way that the function behaves for large values of $|\xi|$.
The characteristic exponents of the processes listed in Table \ref{tab:Charexp} will take high negative values which will dominate $\Phi_{\mathrm{c}}(\xi,s)$ so that as $|\xi|\rightarrow\infty$ we can approximate $s-\psi(\xi)\sim -\psi(\xi)$.
The function to be decomposed in the factorisation stage is therefore $\sim \log[-\psi(\xi)]$. This is an increasing function in $|\xi|$, so the bounds 
for the truncation error of the sinc-based Hilbert transform \cite[Theorems 6.4--6.6]{Feng2008} cannot be used. Moreover, if we consider the truncation errors from Eq.~(\ref{eq:HilbSincApproxTrunc}) for positive and negative values of $k$ individually, we obtain two infinite summations that do not converge. However, Table \ref{tab:Charexp} shows that as $|\xi|\rightarrow\infty$ the values of $\psi(\xi)$ and $\psi(-\xi)$ will become increasingly similar. We can exploit this similarity to find a bound by combining the positive and negative truncations: the truncation error of $f(\xi)=\mathcal{H}[ \log \Phi_{\mathrm{c}}(\xi,s)]$ is bounded as
\begin{align}
|f_{\Delta\xi}(\xi)-f_{\Delta\xi,M}(\xi)|&\! < \Delta\xi\! \sum_{k<-M/2}\!\frac{\log \Phi_{\mathrm{c}}(k\Delta\xi,s)}{\pi(\xi-k\Delta\xi)}+ \Delta\xi \sum_{k>M/2}\frac{\log \Phi_{\mathrm{c}}(k\Delta\xi,s)}{\pi(\xi-k\Delta\xi)}\nonumber\\
& = \Delta\xi \sum_{k>M/2}\left(\frac{\log \Phi_{\mathrm{c}}(k\Delta\xi,s)}{\pi(\xi-k\Delta\xi)}+ \frac{\log \Phi_{\mathrm{c}}(-k\Delta\xi,s)}{\pi(\xi+k\Delta\xi)}\right)\nonumber\\
& = \!\frac{\Delta\xi}{\pi}\!\sum_{k>M/2}\!\frac{\xi\big(\log \Phi_{\mathrm{c}}(k\Delta\xi,s)\!+\!\log \Phi_{\mathrm{c}}(-k\Delta\xi,s)\big)}{\xi^{2}-k^{2} \Delta\xi^{\,2}} \nonumber\\
&\quad +\!\frac{\Delta\xi}{\pi}\!\sum_{k>M/2}\!\frac{k\Delta\xi \big(\log \Phi_{\mathrm{c}}(k\Delta\xi,s)\!-\!\log \Phi_{\mathrm{c}}(-k\Delta\xi,s)\big)}{\xi^{2}-k^{2} \Delta\xi^{\,2}}, \label{eq:errorsum}
\end{align}
where $f_{\Delta\xi}(\xi)$ is the value of the infinite summation in Eq.~(\ref{eq:HilbSincApprox}) and $f_{\Delta\xi,M}(\xi)$ is the result of the truncated summation in Eq.~(\ref{eq:HilbSincApproxTrunc}).

The next step in bounding the error convergence is to show that the expression in Eq.~(\ref{eq:errorsum}) is dominated by the first sum as $M\rightarrow\infty$. As $\psi(k\Delta\xi) \sim \psi(-k\Delta\xi)$ for $k\to\infty$, $\log \Phi_{\mathrm{c}}(k\Delta\xi,s)\!-\!\log \Phi_{\mathrm{c}}(-k\Delta\xi,s)\rightarrow0$ as $k\rightarrow\infty$. However, $k\Delta\xi$ is also present in the numerator and increases linearly with $k$. By determining the rate of decrease of $\log \Phi_{\mathrm{c}}(k\Delta\xi,s)\!-\!\log \Phi_{\mathrm{c}}(-k\Delta\xi,s)$, we show that the second term is bounded as $O(1/k^2)$ and therefore the first term dominates Eq.~(\ref{eq:errorsum}). We then calculate a bound for the error based on the first summation term in Eq.~(\ref{eq:errorsum}). These steps are carried out in a slightly different way depending on the form of the characteristic exponents shown in Table \ref{tab:Charexp}.

\subsubsection{Normal, Merton and Kou processes}\label{sec:error_fact_norm}

For the normal, Merton and Kou processes, when $k\rightarrow\infty$, $\Phi_\mathrm{c}(k\Delta\xi)$ becomes dominated by $\sigma(k\Delta\xi)^2-i\mu k\Delta\xi$ as shown in Table \ref{tab:Charexp}. The parameters $\mu$ and $\sigma$ are specific to the distribution. We can therefore approximate the second expression in the summation by
\begin{align}
\frac{k\Delta\xi \big(\log \Phi_{\mathrm{c}}(k\Delta\xi,s)-\log \Phi_{\mathrm{c}}(-k\Delta\xi,s)\big)}{\xi^2 -k^2 \Delta\xi^2}
&= \frac{k\Delta\xi}{\xi^2 -k^2 \Delta\xi^2} \log \frac{\Phi_{\mathrm{c}}(k\Delta\xi,s)}{\Phi_{\mathrm{c}}(-k\Delta\xi,s)}\nonumber\\
&\sim\frac{k\Delta\xi}{{\xi^2-k^2 \Delta\xi^2}}\log\frac{\sigma^2(k\Delta\xi)^2/2+i\mu(k\Delta\xi)}{\sigma^2(k\Delta\xi)^2/2 -i\mu(k\Delta\xi)}\nonumber\\
&= \frac{k\Delta\xi}{{\xi^2-k^2 \Delta\xi^2}}\log\frac{1+2i\mu/(\sigma^2 k\Delta\xi)}{1-2i\mu/(\sigma^2 k\Delta\xi)}.\label{eq:errorapprox1}
\end{align}
The logarithm in Eq.~(\ref{eq:errorapprox1}) is of the form $\log\frac{1+x}{1-x}$ where $x=\frac{2i\mu}{\sigma^2k\Delta \xi}$. For $x\rightarrow0,\ \log\frac{1+x}{1-x}\sim2x$, thus 
\begin{align}
\frac{k\Delta\xi}{{\xi^{2}-k^{2} \Delta\xi ^{2}}}\log\frac{1+2i\mu/(\sigma^2 k\Delta\xi)}{1-2i\mu/(\sigma^2 k\Delta\xi)}&\sim\frac{k\Delta\xi}{{\xi^{2}-k^{2} \Delta\xi ^{2}}}\frac{4i\mu}{\sigma^2k\Delta \xi}=\frac{4i\mu}{\sigma^2({\xi^{2}-k^{2} \Delta\xi ^{2}})}
\end{align}
gives an approximation for the second term in Eq.~(\ref{eq:errorsum}). Due to the denominator, this is $O(1/k^2)$. Thus, as $\log \Phi_{\mathrm{c}}(k\Delta\xi,s)\!+\!\log \Phi_{\mathrm{c}}(-k\Delta\xi,s)$ is increasing in $k$, the error is indeed dominated by the first term in Eq.~(\ref{eq:errorsum}).

For the normal, Kou and Merton processes, $\Phi_{\mathrm{c}}(k\Delta\xi,s)$ and $\Phi_{\mathrm{c}}(-k\Delta\xi,s)\rightarrow2\log|k\Delta\xi|$ as $k\!\rightarrow\!\infty$. Therefore, the error bound is
\begin{align}
\label{error}
|f_{\Delta\xi}(\xi)-f_{\Delta\xi,M}(\xi)| &< \frac{c\Delta\xi}{\pi}\sum_{k>M/2}\frac{\log\Phi_{\mathrm{c}}(k\Delta\xi,s)}{\xi^{2}-k^{2} \Delta\xi ^{2}},\nonumber\\
& < c_\mathrm{1} \Delta\xi \sum_{k>M/2}\frac{\log(k^{2} \Delta\xi ^{2})}{k^{2} \Delta\xi ^{2}},
\end{align}
where $c$ and $c_\mathrm{1}$ are some constants. Here, as Eq.~(\ref{error}) gives the error at fixed values of $\xi$, i.e.\ the chosen grid points, the $\xi$ can be absorbed into $c$. However, as $M$ increases, our range of $\xi$ values increases. Therefore, as there is a linear dependence of the error bound on $\xi$, we should consider the effect of errors at large values of $\xi$ on the error of the final solution. In doing this we can take account of the shape of the output from the factorisation $\Phi_{\mathrm{c}\pm}(\xi,s)$ which decays as $|\xi|\rightarrow\infty$ and the use of filtering on the input to the next step as described in Section \ref{sec:errperf_decomp}. These effects combine to mean that the error as a proportion of the value of $\Phi_{\mathrm{c}\pm}(\xi,s)$ at high $|\xi|$ is less significant to the error of the overall solution than the relationship between the value of $M$ and the error in $\Phi_{\mathrm{c}\pm}(\xi,s)$ for low values of $|\xi|$.
Approximating the summation by an integral with $M'=M/2$, we obtain
\begin{align}
|f_{\Delta\xi}(\xi)-f_{\Delta\xi,M'}(\xi)| &<c_\mathrm{2}\int^{+\infty}_{M' \Delta\xi}\frac{\log{\xi'}}{{\xi'^{2}}}d\xi' \nonumber\\
& = c_\mathrm{2} \left[\frac{\log \xi'}{\xi'} +\frac{1}{\xi'}\right] _{+\infty}^{M'\Delta\xi}\nonumber\\
& = c_\mathrm{2}\left[\frac{\log M' \Delta\xi}{M' \Delta\xi} +\frac{1}{M' \Delta\xi}\right]\label{eq:errorintegral}
\end{align}
where $c_\mathrm{2}$ is some constant.
Having applied the sinc-based discrete Hilbert transform we can calculate the positive and negative functions using the Plemelj-Sokhotsky relations and then exponentiate the results to obtain the Wiener-Hopf factors. Therefore, using the expression in Eq.~(\ref{eq:errorintegral}), we can bound the truncation error of the Wiener-Hopf factors, and by extension the total error as the truncation error dominates, as
\begin{equation}
\left|\frac{\Phi_{\Delta\xi,\mathrm{c}\pm}(\xi)-\Phi_{\Delta\xi,M',\mathrm{c}\pm}(\xi)}{\Phi_{\Delta\xi,\mathrm{c}\pm}(\xi)}\right|<\left|1-(eM'\Delta\xi)^\frac{\kappa}{M'\Delta\xi}\right|,\label{eq:facterrest}
\end{equation}
where $\kappa$ is some constant. Here, $\Phi_{\Delta\xi,\mathrm{c}\pm}(\xi)$ denotes the (theoretical) Wiener-Hopf factors generated using the series in Eq.~(\ref{eq:HilbSincApprox}) and $\Phi_{\Delta\xi,M',\mathrm{c}\pm}(\xi)$ denotes the Wiener-Hopf factors calculated using the truncated summation in Eq.~(\ref{eq:HilbSincApproxTrunc}).

\begin{figure}
\begin{center}
\includegraphics[width=\textwidth]{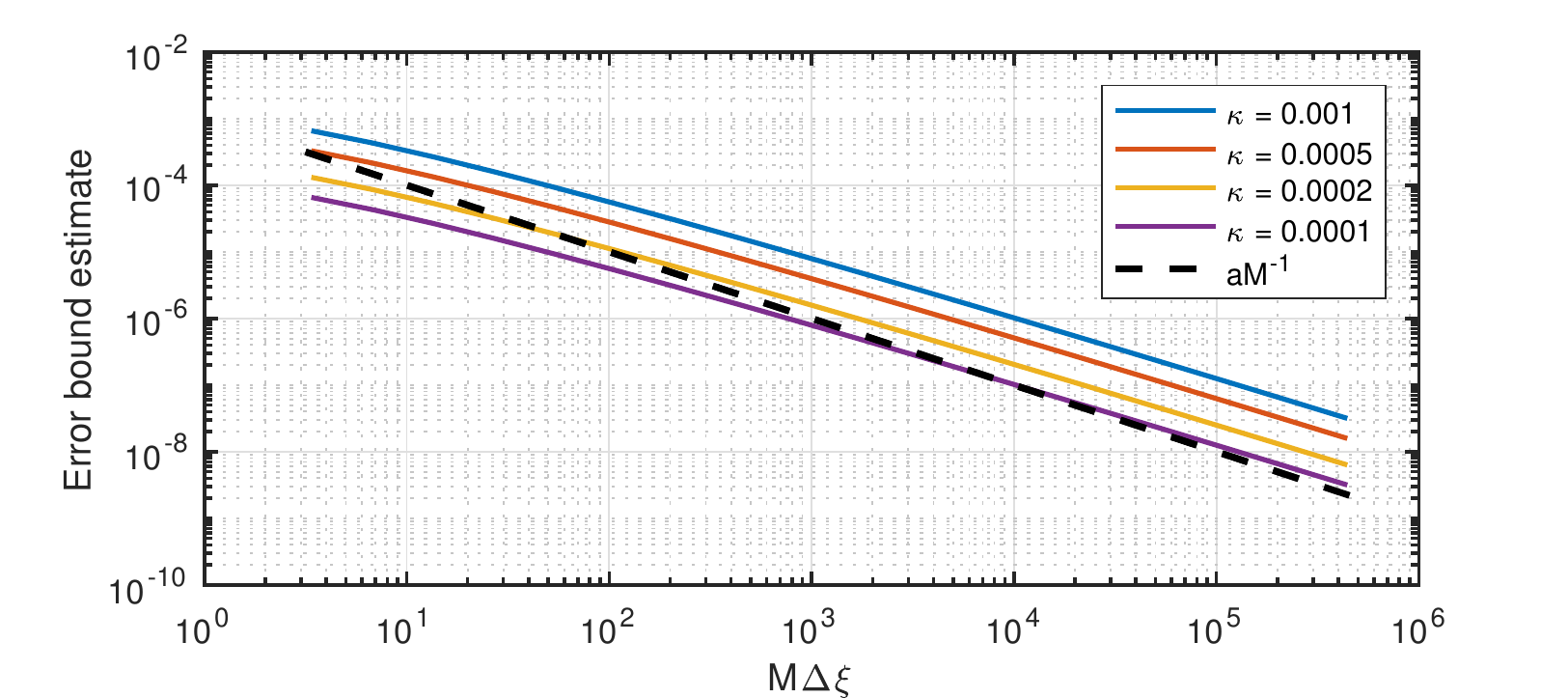}
\caption{Eq.~(\ref{eq:facterrest}) plotted for different values of $\kappa$ to show the estimate of the error bound on the sinc-based numerical factorisation of $\Phi_{\mathrm{c}}(\xi,s)$. Notice that the predicted error bound from the factorisation has a decay that increases in slope as $M\Delta\xi$ increases and is slightly shallower than $O(1/M)$ for the values of $M$ which we are using. Sections \ref{sec:error_fact_norm}, \ref{sec:error_fact_NIG} and \ref{sec:error_fact_VG} show that this bound applies for the normal, NIG, Kou, Merton and VG processes.}
\label{fig:errboundest}
\end{center}
\end{figure}

Figure \ref{fig:errboundest} shows Eq.~(\ref{eq:facterrest}) plotted against $M\Delta\xi$ for different values of $\kappa$. This demonstrates that the predicted error bound from the factorisation has a decay that increases in slope as $M\Delta\xi$ increases and is slightly shallower than $O(1/M)$ for the values of $M$ which we are using.

\subsubsection{Normal inverse Gaussian process}\label{sec:error_fact_NIG}

In the case of the NIG process the characteristic exponent is
\begin{equation}
\psi(\xi)=\delta\big(\sqrt{\alpha^2-(\beta+i\xi)^2}-\sqrt{\alpha^2-\beta^2}\big).
\end{equation}
The presence of a square root around the $i\xi$ and $\xi^{2}$ terms means that as $|k|\!\rightarrow\!\infty$, the equivalent expression to the logarithm in Eq.~(\ref{eq:errorapprox1}) is $\frac{1}{2}\log\frac{1+2i\beta/(k\Delta\xi )}{1-2i\beta/(k\Delta\xi )}$. Furthermore, $\Phi_{\mathrm{c}}(k\Delta\xi,s)$ and $\Phi_{\mathrm{c}}(-k\Delta\xi,s)$ become dominated by $\log|k\Delta\xi|$ as $|k|\rightarrow\infty$. Therefore the only difference between the truncation error bound for the NIG process and the result in Eq.~(\ref{eq:errorintegral}) is the size of the constants.

\subsubsection{Variance gamma process}\label{sec:error_fact_VG}

The characteristic function of the VG process is
\begin{equation}
\psi(\xi) = -\frac{1}{\nu}\log\left(1-i\xi\theta\nu+\frac{1}{2}\nu\sigma^2\xi^2\right).
\end{equation}
This is significantly different from the other characteristic exponents that we have considered, being the log of a polynomial. Similarly to the previous methods, we show that as $k\rightarrow\infty$ the decrease rate of $\log \frac{\Phi_{\mathrm{c}}(k\Delta\xi,s)}{\Phi_{\mathrm{c}}(-k\Delta\xi,s)}$ is at least $O(1/k)$:
\begin{align}
\log \frac{\Phi_{\mathrm{c}}(k\Delta\xi,s)}{\Phi_{\mathrm{c}}(-k\Delta\xi,s)}&\sim\log\frac{\log(-ik\Delta\xi\theta\nu+\nu\sigma^2(k\Delta\xi)^2/2)}{\log\left(ik\Delta\xi\theta\nu+\nu\sigma^2(k\Delta\xi)^2/2\right)}\nonumber\\
&=\log\frac{\log\left(1-\frac{2ik\Delta\xi\theta\nu}{\nu\sigma^2(k\Delta\xi)^2}\right)+\log\frac{ \nu\sigma^2(k\Delta\xi)^2}{2}}{\log\left(1+\frac{2ik\Delta\xi\theta\nu}{\nu\sigma^2(k\Delta\xi)^2}\right)+\log\frac{ \nu\sigma^2(k\Delta\xi)^2}{2}}\nonumber\\
&\sim\log\frac{-\frac{2ik\Delta\xi\theta\nu}{\nu\sigma^2(k\Delta\xi)^2}+\log\frac{\nu\sigma^2(k\Delta\xi)^2}{2}}{\frac{2ik\Delta\xi\theta\nu}{\nu\sigma^2(k\Delta\xi)^2}+\log\frac{\nu\sigma^2(k\Delta\xi)^2}{2}}\nonumber\\
&=\log\frac{1-\frac{2i\theta}{\sigma^2k\Delta\xi\log\left( \nu\sigma^2(k\Delta\xi)^2\right/2)}}{1+\frac{2i \theta}{\sigma^2k\Delta\xi\log\left( \nu\sigma^2(k\Delta\xi)^2/2\right)}}\nonumber\\
&\sim\frac{-4i \theta}{\sigma^2 k\Delta \xi\log\left( \nu\sigma^2(k\Delta\xi)^2/2\right)}.
\end{align}
This decreases quicker than $O(1/k)$ and thus Eq.~(\ref{eq:errorsum}) is dominated by the first term. The equivalent expression to Eq.~(\ref{error}) for the VG process is
\begin{align}
|f_{\Delta\xi}-f_{\Delta\xi,M}| &< \frac{c\Delta\xi}{\pi}\sum_{k>M/2}\frac{\log\Phi_{\mathrm{c}}(k\Delta\xi,s)}{\xi^2-k^2\Delta\xi^2}\nonumber\\
& < c_\mathrm{1} \Delta\xi \sum_{k>M/2}\frac{\log\log(k^2 \Delta\xi^2)}{k^2 \Delta\xi^2}.
\end{align}
As $|\log(\log x)|$ is bounded by $|\log x|$ as $x\rightarrow\infty$, the factorisation error of the method with the VG process is also bounded by Eq.~(\ref{eq:facterrest}).

\subsection{Decomposition error}\label{sec:errperf_decomp}

\begin{figure}
\begin{center}
\includegraphics[width=\textwidth]{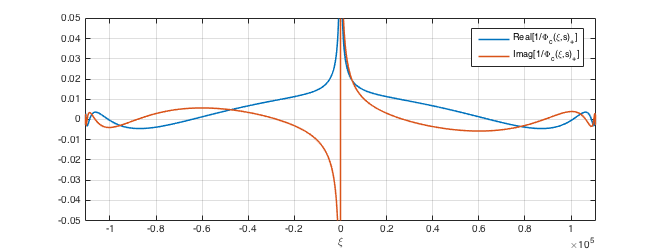}
\caption{Example plot of the real and imaginary parts of $\Phi_{\mathrm{c}+}(\xi,s)$ plotted against $\xi$ with $s = A/(2t)$, as specified for the Abate and Whitt inverse Laplace transform. Notice that although the value of $|\Phi_{\mathrm{c}+}(\xi,s)|$ is bounded by a constant as $|\xi|\rightarrow\infty$, the rate of decay is very slow and we have not been able to determine a decreasing bound.}
\label{fig:factoutput}
\end{center}
\end{figure}

The output of the factorisation is shown in Figure \ref{fig:factoutput}.
The next step in the calculation is to find the positive part with respect to $l$ of $P_{\mathrm{c}}(\xi,s)=\frac{1}{\Phi_{\mathrm{c-}}(\xi,s)}$. We can attempt to bound the truncation error of this calculation by combining the errors from the positive and negative truncations as before:
\begin{align}
\left|f_{\Delta\xi}(\xi)-f_{\Delta\xi,M}(\xi)\right| &=\frac{\Delta\xi}{\pi}\left|\sum_{k>M/2}\frac{P_{\mathrm{c}}(k\Delta\xi)}{\xi-k\Delta\xi}+\sum_{k<-M/2}\frac{P_{\mathrm{c}}(k\Delta\xi)}{\xi-k\Delta\xi}\right|\nonumber\\
& = \frac{\Delta\xi}{\pi}\left|\sum_{k>M/2}\frac{P_{\mathrm{c}}(k\Delta\xi)}{\xi-k\Delta\xi}+\frac{P_{\mathrm{c}}(-k\Delta\xi)}{\xi+k\Delta\xi}\right|\nonumber\\
& = \frac{\Delta\xi}{\pi}\left|\sum_{k>M/2}\frac{\xi[P_{\mathrm{c}}(k\Delta\xi)+P_{\mathrm{c}}(-k\Delta\xi)]}{\xi^2-(k\Delta\xi)^2}+\frac{k\Delta\xi[P_{\mathrm{c}}(k\Delta\xi)-P_{\mathrm{c}}(-k\Delta\xi)]}{\xi^2-(k\Delta\xi)^2}\right|.\label{eq:decomperr}
\end{align}
Figure \ref{fig:factoutput} shows that for high $|\xi|$, $|P_{\mathrm{c}}(\xi)-P_{\mathrm{c}}(-\xi)|$ is bounded from above by a constant. However, we do not have a decreasing bound for $|P_{\mathrm{c}}(\xi)-P_{\mathrm{c}}(-\xi)|$. Therefore we can only bound the second term in Eq.~(\ref{eq:decomperr}) as
\begin{align}
\label{}
&\sum_{k>M/2} \frac{k\Delta\xi[P_{\mathrm{c}}(k\Delta\xi)-P_{\mathrm{c}}(-k\Delta\xi)]}{\xi^2-(k\Delta\xi)^2}< c\sum_{k>M/2} \frac{k\Delta\xi}{\xi^2-(k\Delta\xi)^2}\label{eq:decomperr2}
\end{align}
where $c$ is some positive constant; this does not converge. We can compare it with the discretely monitored version from \cite{Fusai2016} where the first date is taken out of the scheme, meaning that the function undergoing decomposition is multiplied with the characteristic function. For processes other than VG, this restores the exponential slope of the function for high values of $\xi$ which again means that the truncation error of the sinc-based discrete Hilbert transform is exponentially bounded. To improve the error of the decomposition in the continuously monitored case we can improve the slope of the function on the input to the Hilbert transform by using a spectral filter. We use an exponential filter which has previously achieved good results in option pricing applications \citep{ruijter2015application,Phelan2017}. The filter is described by Eq.~(\ref{eq:expFilt}) and its shape is shown in Figure \ref{fig:expfilt2}. Numerical tests have shown that the use of this filter improves the error of the decomposition step so that it no longer limits the error convergence of the pricing scheme. However, the overall error of the pricing procedure will be continue to be limited by the error from the initial factorisation step as described in Eq.~(\ref{eq:facterrest}) and shown in Figure \ref{fig:errboundest}.
\FloatBarrier

\section{Results}\label{sec:Res}

We present results for the Spitzer-Laplace pricing procedure for continuously monitored single and double-barrier options for the NIG, Kou and VG processes. We also show that the error convergence represents a limiting case of the performance of the FGM method for discretely monitored options as $N\rightarrow\infty$ and $\Delta t\rightarrow0$, where $N$ is the number of monitoring dates and $\Delta t$ is the time interval between them.

\subsection{Results for Spitzer-Laplace method for continuously monitored options}\label{sec:Res_Cont}

\begin{figure}[h]
\begin{center}
\includegraphics[width=\textwidth]{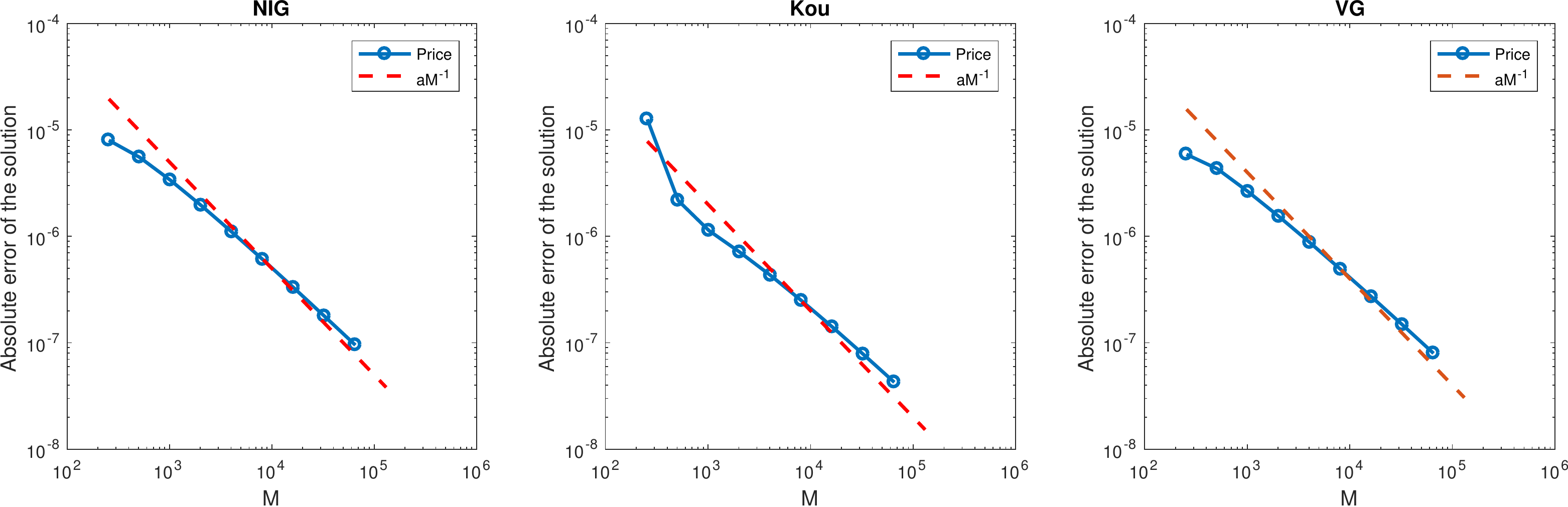}
\caption{Error convergence for a continuously monitored single-barrier option.The error convergence conforms to the calculated error bound (only the first point of the Kou process deviates slightly from the overall behaviour) and shows the typical sub-polynomial error convergence for higher values of $M$.}
\label{fig:N=cont_U=trunc_L=0_80_All}
\end{center}
\end{figure}

\begin{figure}[h]
\begin{center}
\includegraphics[width=\textwidth]{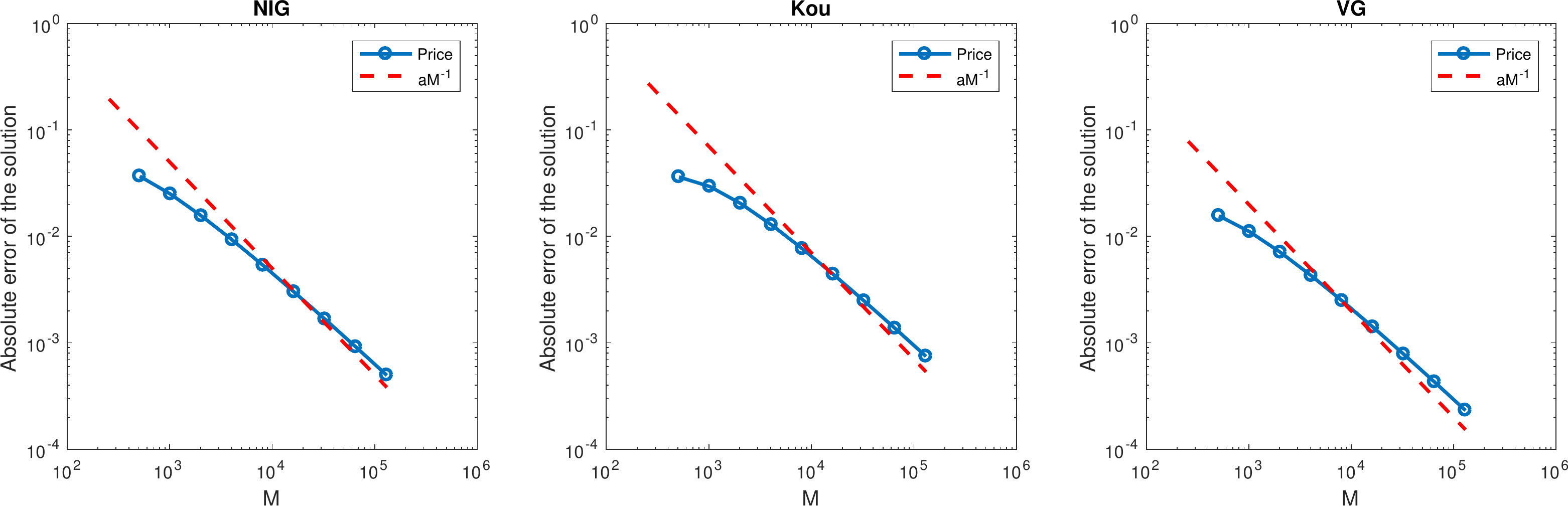}
\caption{Error convergence for the continuously monitored double-barrier option. The absolute error is worse than that for the single-barrier option but the error convergence conforms to the calculated error bound.}
\label{fig:N=cont_U=1_40_L=0_60_All}
\end{center}
\end{figure}

The error convergence for single-barrier down-and-out options is shown in Figure \ref{fig:N=cont_U=trunc_L=0_80_All}. Figure \ref{fig:N=cont_U=1_40_L=0_60_All} shows the results for double-barrier options. The computed prices are given in Tables \ref{tab:Singres} and \ref{tab:Doubres} for single and double-barrier options respectively. Although the absolute error is worse for double-barrier options, the speed of error convergence is very similar for all cases and is slightly worse than $O(1/M)$, which concurs with the simulated results for the factorisation error shown in Figure \ref{fig:errboundest}. The details of the contract and underlying processes are shown in Table \ref{tab:Parasetup} in Appendix A.

\begin{table}[h]
\centering
\begin{tabular}{ll}
\hline\hline
Process & Price \\
\hline
Normal inverse Gaussian & 4.77403523401E-2\\
Kou & 4.32042632202E-2\\
Variance gamma & 4.70627023105E-2\\
\hline\hline
\end{tabular}
\caption{Prices calculated for single-barrier options with the contract details and process parameters described in Table \ref{tab:Parasetup} in Appendix A and $M=2^{17}$.}
\label{tab:Singres}
\end{table}

\begin{table}[h]
\centering
\begin{tabular}{ll}
\hline\hline
Process & Price\\
\hline
Normal inverse Gaussian & 2.78787488E-2\\
Kou & 3.30368034E-2\\
Variance gamma & 2.82666693E-2\\
\hline\hline
\end{tabular}
\caption{Prices calculated for double-barrier options with the contract details and process parameters described in Table \ref{tab:Parasetup} in Appendix A and $M=2^{17}$.}
\label{tab:Doubres}
\end{table}
\FloatBarrier

\subsection{Comparison with the error convergence of Spitzer-$z$ pricing method for discretely monitored options}

In Section \ref{sec:back_spitz_rel} we described the relationship between the Fourier-Laplace based method for continuously monitored options and the FGM method, based in the Fourier-$z$ domain, for discretely monitored options. The latter method, as measured for a single barrier in \cite{Fusai2016} and double barriers in \cite{Phelan2017}, with the number of monitoring dates up to $N\approx10^3$, is exponentially convergent with the number of grid points for the NIG and Kou processes and better than second order polynomially convergent for the VG process. 
Therefore we investigate the performance of the discretely monitored method with a very high number of dates (i.e.~$\Delta t\rightarrow0$), to better understand the difference in performance between the two pricing schemes.

In \cite{Green2010} the error between the discretely and continuously monitored prices was shown to be bounded as $O(1/\sqrt{N})$, where $N$ is the number of monitoring dates.
We therefore also consider whether lower errors might be achieved by approximating the price for a continuously monitored option with the price for a discretely monitored option with a very high number of monitoring dates.

\begin{figure}
\begin{center}
\includegraphics[width=\textwidth]{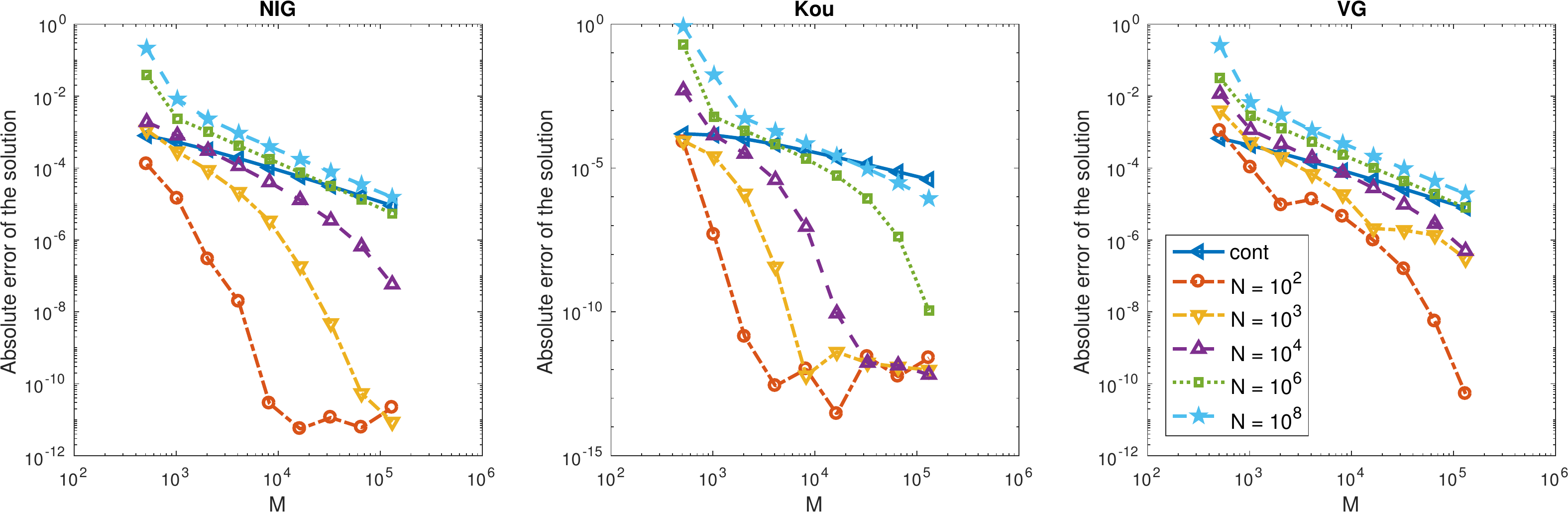}
\caption{Error as a function of the grid size $M$ for continuously monitored single-barrier options compared to discretely monitored options as the number of monitoring dates $N$ increases. The error for each number of dates (including continuous monitoring) is calculated against the price for the same number of dates with $2^{18}$ grid points. For all processes, as $\Delta t\rightarrow 0$ the slope of the error convergence of the discretely monitored scheme approaches that of the continuously monitored scheme.}
\label{fig:NdatesvErr_U=trunc_L=0_80_All}
\end{center}
\end{figure}

\begin{figure}
\begin{center}
\includegraphics[width=\textwidth]{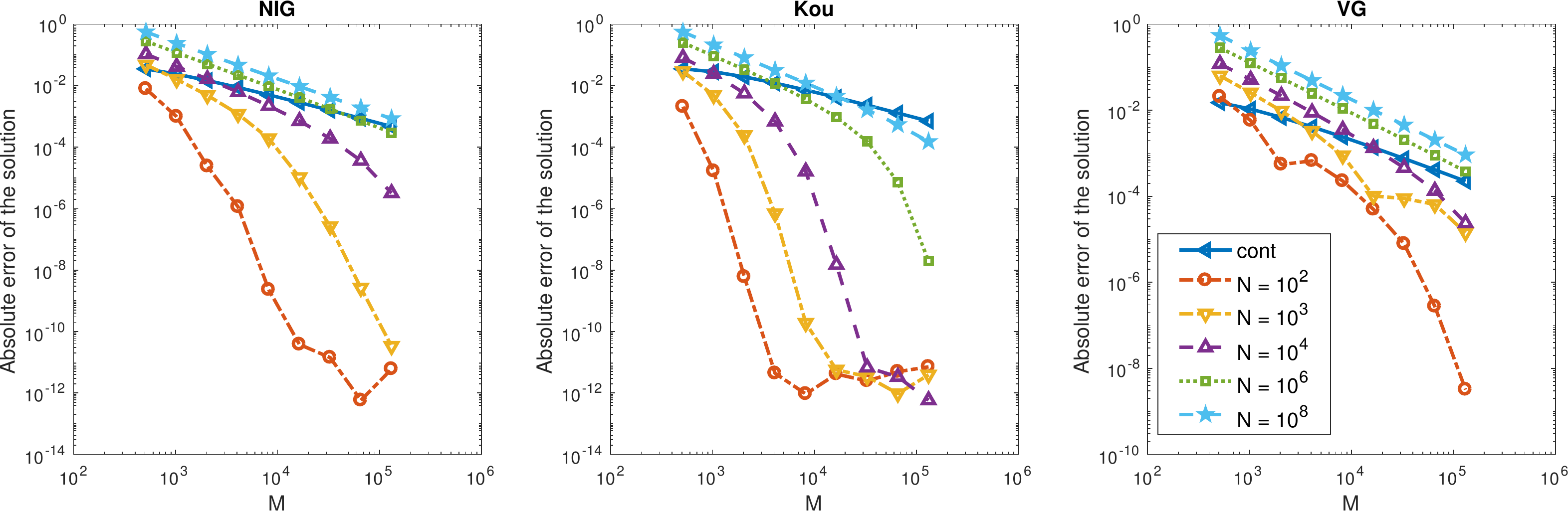}
\caption{Error as a function of the grid size $M$ for continuously monitored double-barrier options compared to discretely monitored options as the number of monitoring dates $N$ increases. The error for each number of dates (including continuous monitoring) is calculated against the price for the same number of dates with $2^{18}$ grid points. For all processes, as $\Delta t\rightarrow 0$ the slope of the error convergence of the discretely monitored scheme approaches that of the continuously monitored scheme.}
\label{fig:NdatesvErr_U=1_40_L=0_60_All}
\end{center}
\end{figure}

We use the same implementation as the one described in \cite{Fusai2016} for single-barrier options and \cite{Phelan2017} for double-barrier options, although the maximum number of monitoring dates is far higher than would ever be used for discretely monitored options in practice. Due to the $O(1/\sqrt{N})$ error bound between the prices for continuously and discretely monitored options, we must select a very large number of monitoring dates in order for this effect to be less significant than the error from the continuously monitored pricing method. The error convergence of the discrete pricing method as $N\rightarrow\infty$ (or $\Delta t\rightarrow0$) is shown in Figures \ref{fig:NdatesvErr_U=trunc_L=0_80_All} and \ref{fig:NdatesvErr_U=1_40_L=0_60_All}. The results show that as $\Delta t\rightarrow0$, the error convergence for discrete monitoring degrades until it approaches that of continuously monitored options. Moreover, it demonstrates that, rather than being an anomaly, the error convergence of the continuously monitored method is consistent with that of the discretely monitored method as $\Delta t\rightarrow0$. This can be understood by considering how $\Psi(\xi,\Delta t)$ changes with $\Delta t$ for the discrete example. As $\Delta t\rightarrow 0$, $\Psi(\xi,\Delta t)=e^{\psi(\xi)\Delta t}$ decays to $0$ more and more slowly as $|\xi|\rightarrow\infty$. Therefore the error convergence of the pricing technique for continuously monitored barrier options is a limit of the error convergence for the discrete case as $\Delta t\rightarrow 0$.

\begin{figure}
\begin{center}
\includegraphics[width=\textwidth]{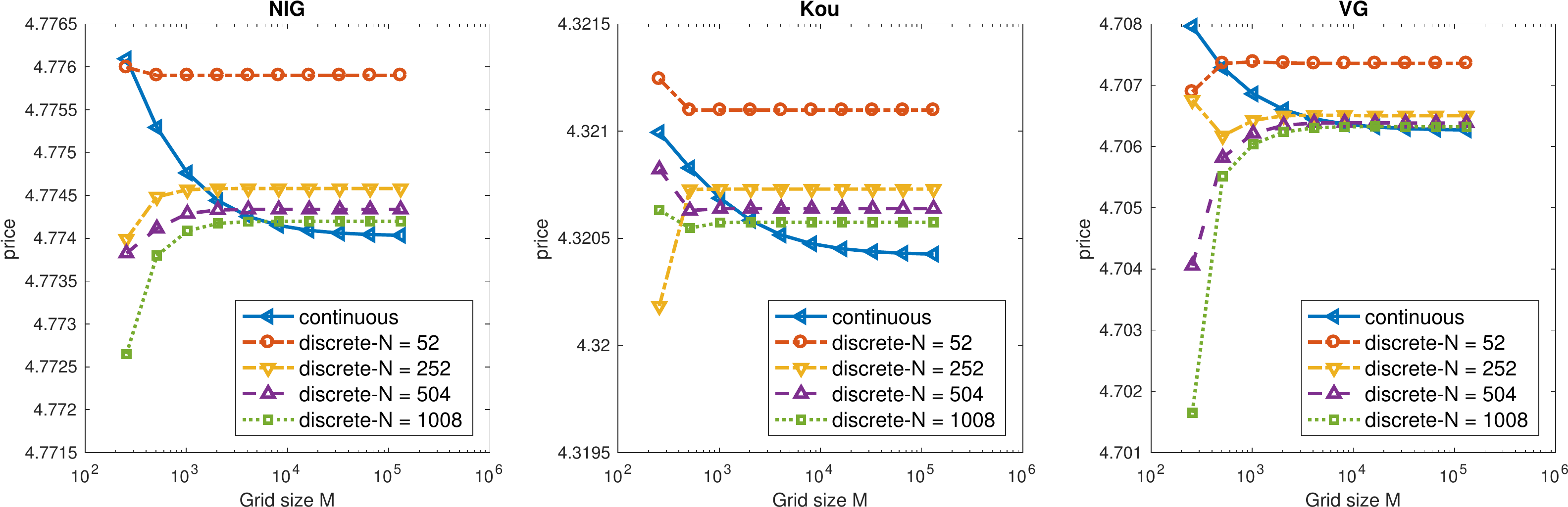}
\caption{Price plotted against the grid size $M$ for continuously monitored single-barrier options compared to discretely monitored options as the number of monitoring dates $N$ increases. Note that the larger the value of $N$, the closer the price is to the continuously monitored option price.}
\label{fig:Ndatesvprice_U=trunc_L=0_80_All}
\end{center}
\end{figure}

\begin{figure}
\begin{center}
\includegraphics[width=\textwidth]{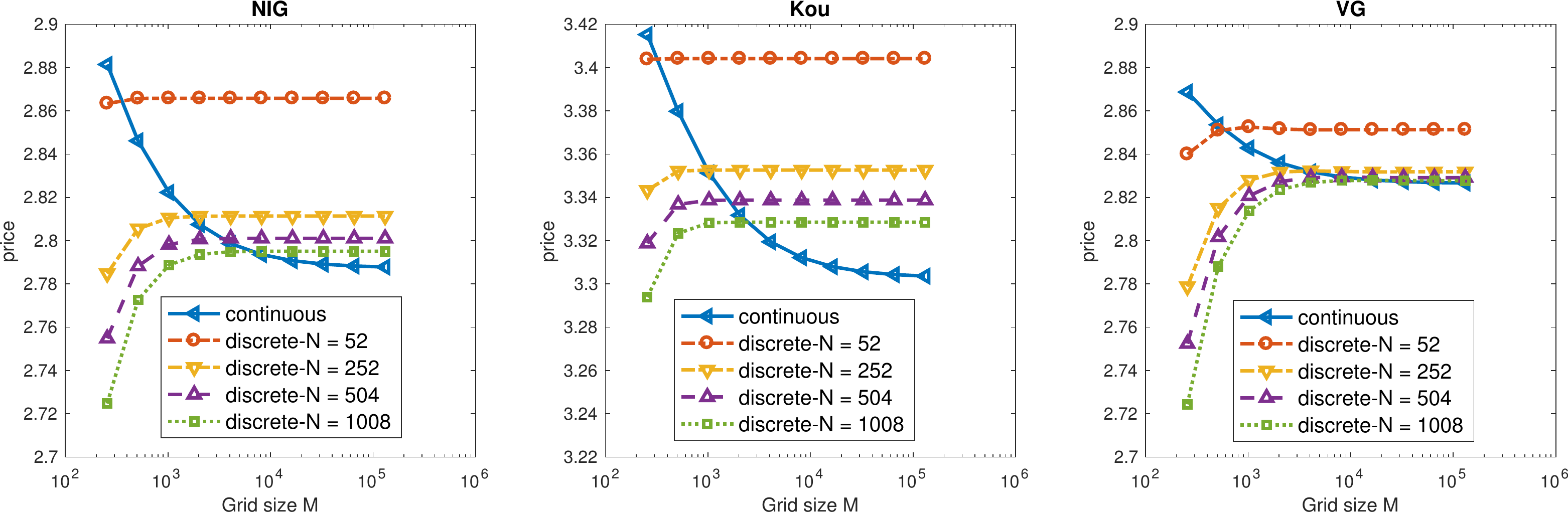}
\caption{Price plotted against grid size $M$ for continuously monitored double-barrier options compared to discretely monitored options as the number of monitoring dates $N$ increases. Note that the larger the value of $N$, the closer the price is to the continuously monitored option price.}
\label{fig:Ndatesvprice_U=1_40_L=0_60_All}
\end{center}
\end{figure}

Computed prices for continuously and discretely monitored options are plotted against $M$ in Figures \ref{fig:Ndatesvprice_U=trunc_L=0_80_All} and \ref{fig:Ndatesvprice_U=1_40_L=0_60_All}. In addition, computation times of the pricing methods for the discrete and continuously monitored methods are shown in Tables \ref{tab:SingerrvCPUcomparison} and \ref{tab:DouberrvCPUcomparison}. Figures \ref{fig:Ndatesvprice_U=trunc_L=0_80_All} and \ref{fig:Ndatesvprice_U=1_40_L=0_60_All} show that, as expected, the larger the number of monitoring dates the closer the price is to the continuously monitored price. However, they also show that the direction of convergence depends on the type of option and the process being used. Therefore, in order to obtain the CPU times in Tables \ref{tab:SingerrvCPUcomparison} and \ref{tab:DouberrvCPUcomparison} we take the lowest time where the convergence error for the discretely monitored method is significantly lower (i.e.\ ten times) than the error compared to the price for the continuously monitored case with maximum $M$. This shows that for relatively high errors, $\approx10^{-4}$ for a single barrier and $\approx10^{-2}$ for double barriers, the discretely monitored method is slightly quicker. However, the discretely monitored method is unable to achieve the lower errors, $\approx10^{-6}$ for a single barrier and $\approx10^{-4}$ for double barriers, which are attained by the continuously monitored method and therefore is not a sufficiently accurate approximation.

\begin{figure}
\begin{center}
\includegraphics[width=0.45\textwidth]{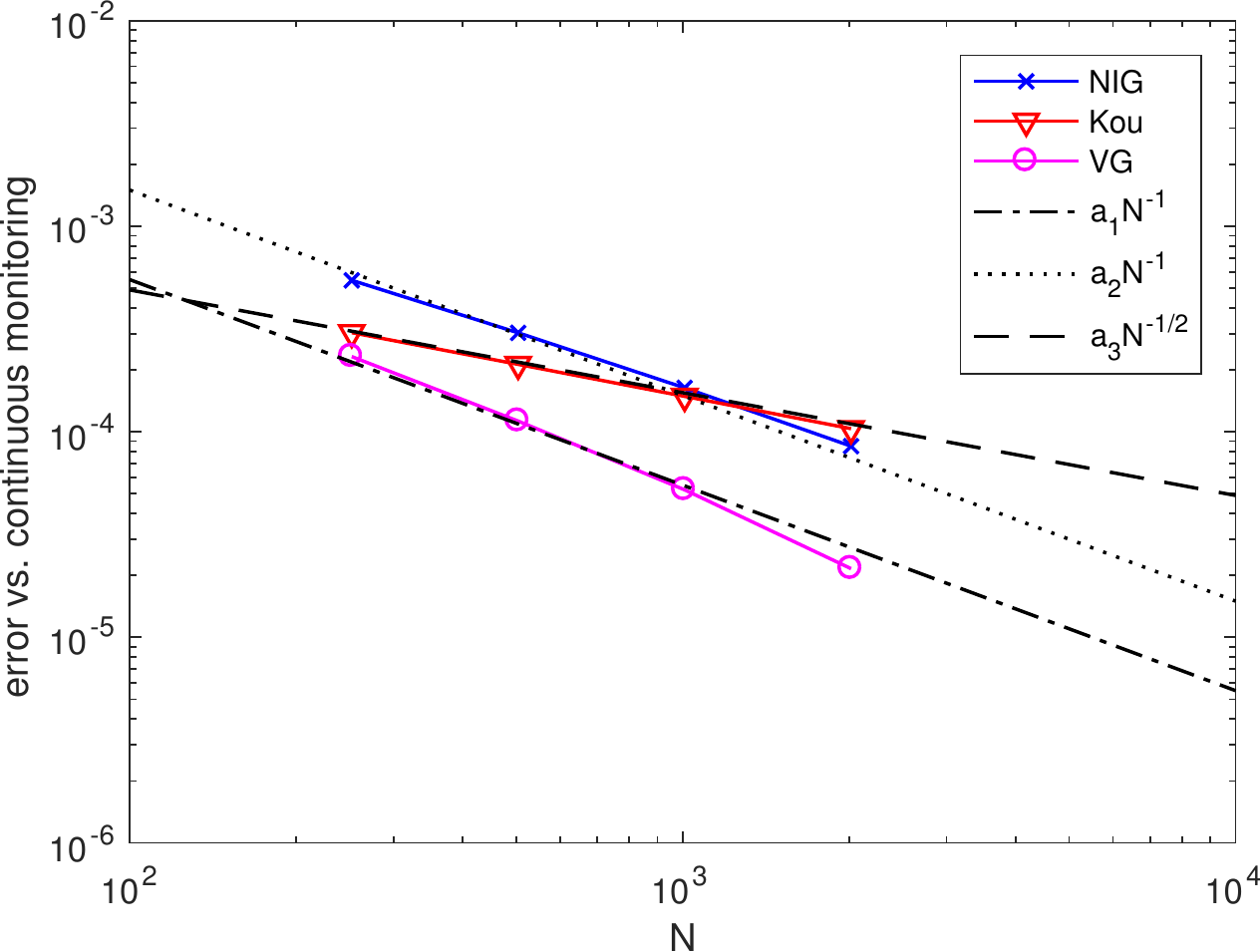}
\includegraphics[width=0.45\textwidth]{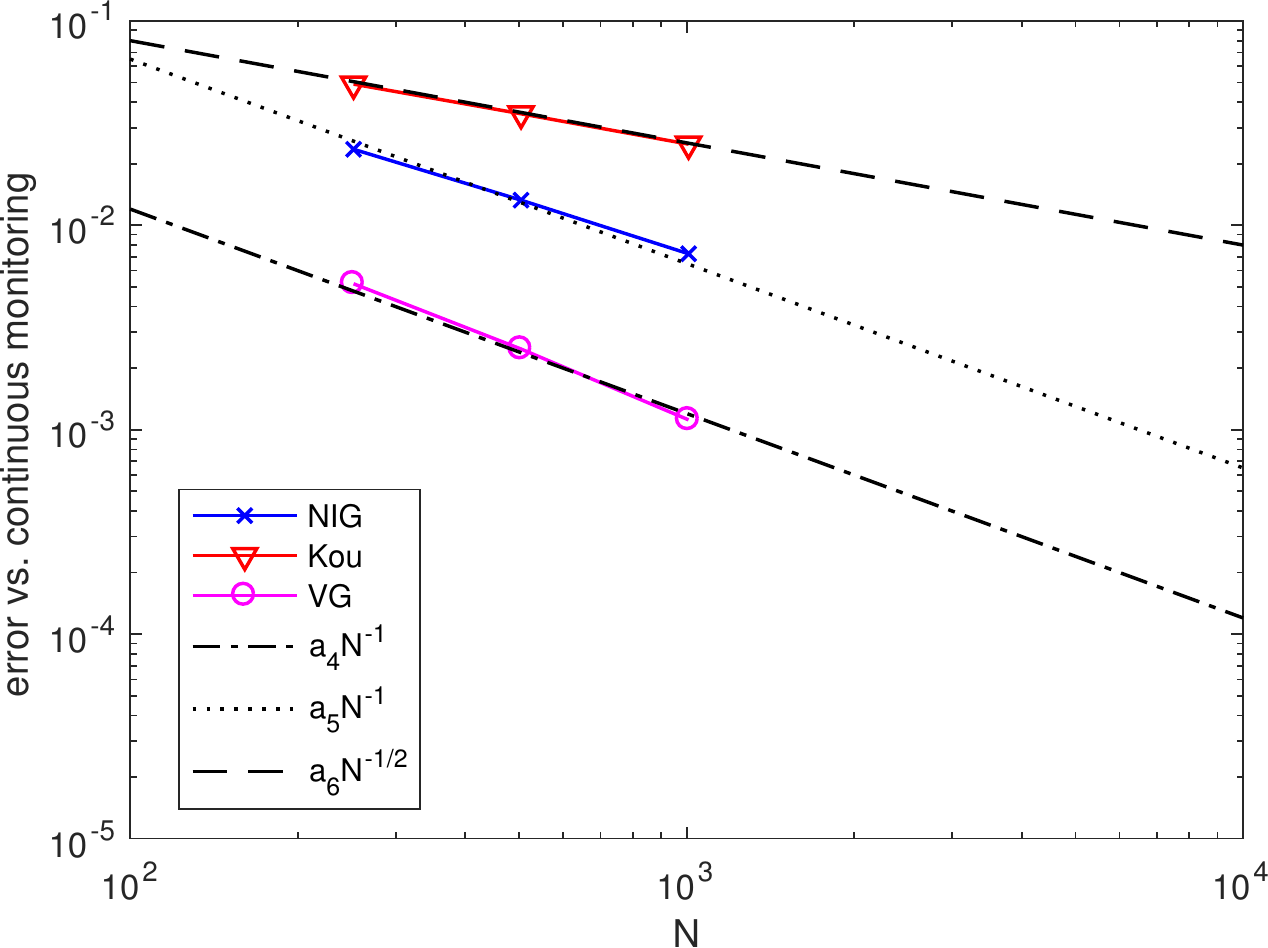}
\caption{Error for discretely monitored barrier options used as an approximation of the price for the continuously monitored case, plotted as a function of the number of monitoring dates. The error is calculated as the difference between the prices for discrete and continuous monitoring at the maximum grid size of $2^{17}$ Results for single and double-barrier options are displayed on the left and right hand plots respectively. Notice that the error for the Kou process converges as $O(1/\sqrt{N})$, whereas the error for the NIG and VG processes converges a rate of approximately $O(1/N)$.}
\label{fig:barrierdatesconvergence}
\end{center}
\end{figure}

Can we can achieve a better approximation of the continuous method by increasing the number of monitoring dates further? Previous literature, e.g.\ \cite{Green2010}, has shown that the convergence of the discrete method to the continuous method with increasing monitoring dates is $O(1/\sqrt{N})$. From Figure \ref{fig:barrierdatesconvergence} we can observe that, although the discrete method with the Kou process does indeed have this rate of convergence, it achieves approximately $O(1/N)$ with the NIG and VG processes. Therefore, if we wished to decrease the error of the discrete approximation so that it is significantly (i.e.\ ten times) less than the continuous case then we would have to increase the maximum number of monitoring dates in Tables \ref{tab:SingerrvCPUcomparison} and \ref{tab:DouberrvCPUcomparison} by 100 times for the NIG and VG processes and by $200^2$ times for the Kou process. We can see from Figures \ref{fig:NdatesvErr_U=trunc_L=0_80_All} and \ref{fig:NdatesvErr_U=1_40_L=0_60_All} that at $M=2^{17}$ the discrete methods with these numbers of dates have an error which is worse than the required accuracy of ten times better than the continuous method. Thus the only possibility would be to also increase $M$, and by extension the CPU time of the discrete monitoring method, causing its computational cost to be greater than that for continuous monitoring.

\begin{table}[h]
\centering
\begin{tabular}{lrllr}
\hline\hline
Process & Dates & Error & Time & $M$\\
\hline and & cont & 3.21E-04 & 0.07 & 1024\\
\cline{2-5}
& 1008 & 1.60E-04 & 0.07 & 4096\\
& cont & 1.86E-04 & 0.13 & 2048\\
\cline{2-5}
& cont & 5.74E-05 & 0.50 & 8192\\
\cline{2-5}
& cont & 1.69E-05 & 2.03& 32768\\
\hline
\multirow{8}{*}{VG} & 252 & 2.31E-04 & 0.04 & 2048\\
& cont & 2.57E-04 & 0.06 & 1024\\
\cline{2-5}
& 504 & 1.13E-04 & 0.11 & 4096\\
& cont & 1.49E-04 & 0.17 & 2048\\
\cline{2-5}
& 1008 & 5.29E-05 & 0.14 & 8192\\
& cont & 4.74E-05 & 0.49 & 8192\\
\cline{2-5}
& cont & 1.43E-05 & 2.01 & 32768\\
\cline{2-5}
& cont & 4.19E-06 & 11.52 & 131072\\
\hline
\multirow{7}{*}{Kou} & 252 & 3.02E-04 & 0.01 & 512\\
& cont & 1.57E-04 & 0.02& 256\\
\cline{2-5}
& 504 & 2.03E-04 & 0.01 & 512\\
& cont & 1.57E-04 & 0.02& 256\\
\cline{2-5}
& 1008 & 1.47E-04 & 0.02 & 1024\\
& cont & 1.57E-04 & 0.02& 256\\
\cline{2-5}
& cont & 6.69E-05 & 0.12 & 2048\\
\cline{2-5}
& cont & 2.35E-05 & 0.49 & 8192\\
\cline{2-5}
& cont & 7.36E-06 & 2.07 & 32768\\
\hline\hline
\end{tabular}
\caption{CPU times and errors for the continuously monitored method and the discretely monitored method as an approximation to continuous monitoring for the single-barrier case. The CPU times for the discretely monitored price are chosen for the grid size $M$ which gives the lowest CPU time where the convergence error is significantly (about ten times) lower than the error compared to the continuously monitored price.}
\label{tab:SingerrvCPUcomparison}
\end{table}

\begin{table}[h]
\centering
\begin{tabular}{lrllr}
\hline\hline
Process & Dates & Error & Time & $M$\\
\hline
\multirow{8}{*}{NIG} & 252 & 2.35E-02 & 0.16 & 4096 \\
& cont & 2.38E-02 & 0.11 & 512\\
\cline{2-5}
& 504 & 1.32E-02 & 0.14 & 4096\\
& cont & 1.48E-02 & 0.24 & 1024\\
\cline{2-5}
& 1008 & 7.24E-03 & 0.29 & 8192\\
& cont & 8.78E-03 & 0.54 & 2048\\
\cline{2-5}
& cont & 1.58E-03 & 5.03 & 16384\\
\cline{2-5}
& cont & 4.69E-04 & 20.94 & 65536\\
\cline{2-5}
\hline
\multirow{8}{*}{VG} & 252 & 5.10E-03 & 0.08 & 2048\\
& cont & 6.86E-03 & 0.24 & 1024\\
\cline{2-5}
& 504 & 2.51E-03 & 0.13 & 4096\\
& cont & 2.40E-03 & 1.15& 4096\\
\cline{2-5}
& 1008 & 1.15E-03 & 0.29 & 8192\\
& cont & 1.36E-03 & 2.44 & 8192\\
\cline{2-5}
& cont & 7.56E-04 & 5.16 & 16384\\
\cline{2-5}
& cont & 2.25E-04 & 21.21 & 65536\\
\hline
\multirow{8}{*}{Kou} & 252 & 4.90E-02 & 0.03 & 1024\\
& cont & 3.54E-02 & 0.07 & 256\\
\cline{2-5}
& 504 & 3.51E-02 & 0.07 & 2048\\
& cont & 3.54E-02 & 0.07 & 256\\
\cline{2-5}
& 1008 & 2.47E-02 & 0.04 & 1024\\
& cont & 2.84E-02 & 0.14& 512\\
\cline{2-5}
& cont & 7.23E-03 & 1.19 & 4096\\
\cline{2-5}
& cont & 2.33E-03 & 5.30 & 16384\\
\cline{2-5}
& cont & 7.03E-04 & 21.05 & 65536\\
\hline\hline
\end{tabular}
\caption{CPU times and errors for the continuously monitored method and the discretely monitored method as an approximation to continuous monitoring for the double-barrier case. The CPU times for the discretely monitored price are chosen for the grid size $M$ which gives the lowest CPU time where the convergence error is significantly (about ten times) lower than the error compared to the continuously monitored price.}
\label{tab:DouberrvCPUcomparison}
\end{table}

\FloatBarrier
\section{Conclusions}
We showed that the numerical method for calculating the discretely monitored Spitzer identities described by \cite{Fusai2016} can be modified for continuous monitoring by using the Fourier-Laplace domain instead of the Fourier-$z$ domain. We implemented this with the inverse Laplace transform by \cite{Abate1992_2,Abate1995} which achieves an accuracy of approximately $10^{-11}$, sufficient for our chosen application of pricing barrier options. We presented results showing that the conversion from discrete to continuous monitoring means that exponential convergence is no longer achieved, but instead the error convergence becomes sub-polynomial due to the performance of the Wiener-Hopf factorisation.
By examining the effect of truncating the sinc-based discrete Hilbert transform, we were able to provide an error bound which is well matched to the observed accuracy of the pricing procedure for continuously monitored options.

It is notable that previous papers have shown that the discretely monitored case achieves exponential convergence \citep{Fusai2016,Phelan2017} but the continuous case described here does not. However, we showed that, as the number of monitoring dates increases and $\Delta t\rightarrow0$, the error convergence for the discretely monitored case degrades and approaches that of the continuously monitored case. Thus, the performance of the pricing technique for continuously monitored barrier options is consistent with previous results, being a limit of the error convergence for the discretely monitored case.

Furthermore we have compared the error vs.\ computational time of the continuously monitored scheme with that of an approximate solution generated by the discretely monitored scheme with a high number of monitoring dates. We show that, for higher errors, the discrete scheme may produce a rapidly calculated approximation to the continuously monitored scheme, but when lower errors are required the continuously monitored scheme must be used.

\section*{Acknowledgments}
The support of the Economic and Social Research Council (ESRC) in funding the Systemic Risk Centre (grant number ES/K002309/1) and of the Engineering and Physical Sciences Research Council (EPSRC) in funding the UK Centre for Doctoral Training in Financial Computing and Analytics (grant number 1482817) are gratefully acknowledged.

\bibliography{papergeneric_30_11_17}
\bibliographystyle{ormsv080}
\newpage
\section*{Appendix A}
Table \ref{tab:Parasetup} contains all the parameters used for the numerical experiments which produced the results presented in Section \ref{sec:Res}.
\begin{table}[h]
\begin{center}
\begin{tabular}{lllr}
\hline\hline
& Description & Symbol & Value \\
\hline
\multirow{9}{*}{Option parameters}&Maturity & $T$& 1 year \\
&Initial spot price &$S_0$ & 1\\
&Strike &$K$ & 1.1\\
&Upper barrier (double-barrier) & $U$ & 1.40\\
&Lower barrier (double-barrier) & $U$ & 0.60\\
&Upper barrier (down-and-out) & $U$ & $+\infty$ \\
&Lower barrier (down-and-out) & $L$ & 0.80\\
&Risk-free rate &$r$ & 0.05\\
&Dividend rate &$q$& 0.02\\
\hline
Model & $\Psi(\xi,t)$ & Symbol & Value\\
\hline
\multirow{3}{*}{NIG}&\multirow{3}{*}{$e^{-t\left(\sqrt{\alpha^2-(\beta+i\xi)^2}+\sqrt{\alpha^2-\beta^2}\right)}$} & $\alpha$ & 15\\
& & $\beta$ & -5\\
& & $\delta$ & 0.5\\
\hline
\multirow{5}{*}{Kou}&\multirow{5}{*}{$e^{-t\left(\frac{\sigma^2\xi^2}{2}-\lambda\left(\frac{(1-p)\eta_2}{\eta_2+i\xi}+\frac{p\eta_1}{\eta_1-i\xi}-1\right)\right)}$} & $p$ & 0.3\\
& & $\lambda$ & 3\\
& & $\sigma$ & 0.1\\
& &$\eta_1$ & 40\\
& & $\eta_2$ & 12\\
\hline
\multirow{3}{*}{VG}&\multirow{3}{*}{$(1-i\nu\xi\theta+\nu\sigma^2\xi^2/2)^{-t/\nu}$} & $\theta$ & $\frac{1}{9}$\\
\noalign{\vskip 0.5mm}
& & $\sigma$ &$\frac{1}{3\sqrt{3}}$\\
\noalign{\vskip 0.5mm}
& & $\nu$ & 0.25 \\
\hline\hline
\end{tabular}
\end{center}
\caption{Parameters for the numerical tests and processes used; $\Psi(\xi,t)$ is the characteristic function of the process that models the log return of the underlying asset.}
\label{tab:Parasetup}
\end{table}

\end{document}